\documentclass[aps,prl,twocolumn,superscriptaddress,nofootinbib]{revtex4-2}
\usepackage[T1]{fontenc}
\usepackage[utf8]{inputenc}
\usepackage{bm,amsmath,amssymb,graphicx}
\usepackage{microtype}
\usepackage[colorlinks=true,linkcolor=blue,citecolor=blue,urlcolor=blue]{hyperref}
\usepackage{comment}
\usepackage{isotope}
\usepackage{orcidlink}
\usepackage{placeins}

\newcommand{\B}{\mathcal{B}}

\begin{document}

\title{Avoiding Blindness in Baryon Number Violating Processes: \\
Free-Beam and Intranuclear Paths to Neutron-Antineutron Transitions }

\author{Joshua L. Barrow\orcidlink{0000-0002-7319-3339}\,}
\affiliation{The University of Minnesota, Twin Cities}
\author{Peter Fierlinger}
\affiliation{Physik-Department, Technische Universit\"at M\"unchen, 85748 Garching, Germany}
\author{Yuri Kamyshkov}
\affiliation{Department of Physics and Astronomy, University of Tennessee, Knoxville, TN 37996, USA}
\author{Bernhard Meirose\orcidlink{0000-0003-0032-7022}\,}
\affiliation{Institutionen f\"or Fysik, Chalmers Tekniska H\"ogskola, Sweden}
\affiliation{Department of Physics, Lund University, 221\,00 Lund, Sweden}
\author{David Milstead}
\affiliation{Department of Physics, Stockholm University, SE-106 91 Stockholm, Sweden}
\affiliation{Oskar Klein Centre, Stockholm, Sweden}
\author{Rabindra N.~Mohapatra}
\affiliation{Department of Physics, University of Maryland, College Park, MD 20742, USA}
\author{Valentina Santoro}
\affiliation{Department of Physics, Lund University, 221\,00 Lund, Sweden}
\affiliation{European Spallation Source ERIC, 221\,00 Lund, Sweden}

\date{\today}

\begin{abstract}
Experimental searches for neutron--antineutron ($n \rightarrow \bar n$) transitions can be considered via two approaches: conversion in free-neutron beams and intranuclear transformation leading to matter instability in large-mass detectors. Plans for next-generation searches make it timely to highlight the complementarity, necessity, and limitations of each method. Converting the bound neutron limit into one for free neutrons traditionally utilizes nucleus-specific estimates of the in-medium suppression of $n \rightarrow \bar n$, obtained within mean-field theory under a single-operator assumption. This paper highlights how this suppression can be scenario-dependent, which can lead to deviations from the standard approach that can span several orders of magnitude. A further goal of the paper is to point out the need for a broader phenomenology program for $n\rightarrow \bar{n}$ that is akin to those developed for electric dipole moments and other systems for which short-distance new physics must be studied in-medium.
\end{abstract}

\maketitle

\section*{Introduction}
\noindent Of all of the hitherto observed conservation laws, the apparent protection of baryon number ($\mathcal{B}$) is perhaps the most fragile. Theories of baryogenesis~\cite{Riotto:1999yt,Dine:2003ax,Babu:2006psb,Canetti:2012zc,Babu:2013yca} require baryon number violation (BNV) as a Sakharov condition~\cite{Sakharov:1967dj}. Similarly for the conservation of lepton number ($\mathcal{L}$), baryon number ($\mathcal{B}$) is conserved in the Standard Model (SM) at the perturbative level due to an accidental global symmetry~\cite{Weinberg:1979sa,Wilczek:1979hc}. Both conservation laws are typically not respected when the SM is extended. Away from the perturbative sector, the violation of $\mathcal{B}$ and $\mathcal{L}$ is predicted to occur within the electroweak sector of the SM via sphaleron and instanton processes~\cite{tHooft:1976rip,Kuzmin:1985mm} where the quantity ($\mathcal{\mathcal{B}-\mathcal{L}}$) is instead conserved. While $\Delta\B=1$ searches~\cite{Abe:2017pde_pi0, Abe:2014nuK, Hirata:1989leptonMeson} dominated attention in early GUT paradigms~\cite{Georgi:1974gg, Langacker:1981GUTreview, Hebecker:2024PDG_GUTs}, decades of null results motivate complementary paths. Many well-motivated frameworks of physics beyond the SM violate $\mathcal{B}$ by two units~\cite{Mohapatra:1979ia, Mohapatra:1981prd,Nussinov:2002,Babu:2006psb, Barbier:2004ez, Phillips:2014fgb,Calibbi:2016ukt}.

The most sensitive way to search for a $\Delta \mathcal{B}=2$ transition is via neutron-antineutron transformations ($n\rightarrow \bar{n}$). One method to do this considers an antineutron produced in a long free neutron beam~\cite{Fidecaro1985_free,Bressi1990_free,Baldo-Ceolin:1994hzw,Addazi:2020nlz}, while the other utilizes large-volume underground detectors~\cite{Frejus1990_bound,Soudan22002_bound,SNO2017_bound,SuperK2015_bound,SuperK2021_bound} to search for intranuclear transitions which lead to matter instability. Each can lead to a highly unique $\bar{n}N$ annihilation signature of multipion final states. While ``free'' neutron searches are generally inhibited only by energy level splittings $\delta E$ originating from environmental magnetic fields which can be largely shielded away, the strong nuclear field inhibits the $n\rightarrow \bar{n}$ transition. A nucleus-specific 
parameter, $R=T_b/\tau_{f}^2$, is generically applied to any experimentally determined intranuclear (bound) lifetime lower limit ($T_{b}$) to recover the associated square of the oscillation period for free neutrons ($\tau_{f}$). Estimations of $R$ are based on phenomenologically mature mean-field/optical potential models assuming a contact interaction and momentum-independent microscopic mixing~\cite{Alberico:1982nu,Dover:1982wv,Alberico:1984wk,Dover:1985hk,Dover:1989zz,Alberico:1990ij, Hufner:1998gu,Friedman:2008es,Barrow:2022nxb,Barrow:2019viz}. Typical values of $R$ lie in the range of $10^{22}-10^{23}\,$s$^{-1}$. Remarkably, given that sensitivities of free and bound searches are determined by wholly different sets of practical constraints, the limits from each method are very similar. The most competitive free or free-equivalent oscillation time lower limits from each method are $0.86\times 10^{8}\,$s~\cite{Baldo-Ceolin:1994hzw} from a direct search at the Institut Laue Langevin (ILL), and an inferred limit of $4.7 \times 10^8\,$s~\cite{SuperK2021_bound} from Super-Kamiokande. With neither method offering dominant sensitivity, it is important that physics which is not captured in the standard $R$ estimations be considered. 

As will be discussed in this letter, scenarios beyond the mean field paradigm used in standard $R$ determinations may change the intranuclear suppression of $n\rightarrow \bar{n}$ by multiple orders of magnitude. These include multi-operator interference and non-local transitions. 
These are relatively unexplored in the literature,  particularly with respect to possible cancellations among operator contributions and potential nuclear or kinematic enhancements, unlike, for example, $CP$-violating operators underlying electric dipole moment (EDM) observables which are tightly bounded by a broad and complementary suite of measurements (neutron, diamagnetic atoms, paramagnetic molecules, etc.)~\cite{Pospelov:2005pr, Engel:2013lsa,LiuTimmermans:2007PRC, FlambaumGinges:2002PRA,deVries:2011PRC,Bsaisou:2015AOP}. Neutrinoless double-$\beta$ decay~\cite{Engel:2016xgb,Cirigliano:2018Master},  dark-matter–nucleus scattering~\cite{Fitzpatrick:2012ix,Anand:2013yka,Schneck:2015eqa}, and $\mu\!\to\! e$ conversion~\cite{Calibbi:2018RivNuovo,Kitano:2002PRD} provide further examples of systems for which there is an interplay between short distance and nuclear physics. 

Each $n\rightarrow \bar{n}$ search method is complementary in the physics that it can elucidate. The observation of $n\rightarrow \bar{n}$ at a free neutron experiment provides a powerful probe of Lorentz/CPT-violating backgrounds in the Standard-Model Extension (SME) and small Weak Equivalence Principle (WEP) violations~\cite{Babu:2015axa,Addazi:2016rgo} to which bound neutron searches are blind. The related $\Delta \mathcal{B}=2$ process of dinucleon decay is uniquely available via bound nucleon searches~\cite{Super-Kamiokande:2014hie,Super-Kamiokande:2015jbb,Super-Kamiokande:2015pys}. Searches for $n\rightarrow \bar{n}$ in free space and in various mediums may also give sensitivity to different sets of operator projections.  

\subsection{Search status and prospects}
\begin{figure}[t!]
  \centering
    \includegraphics[width=1.\columnwidth]{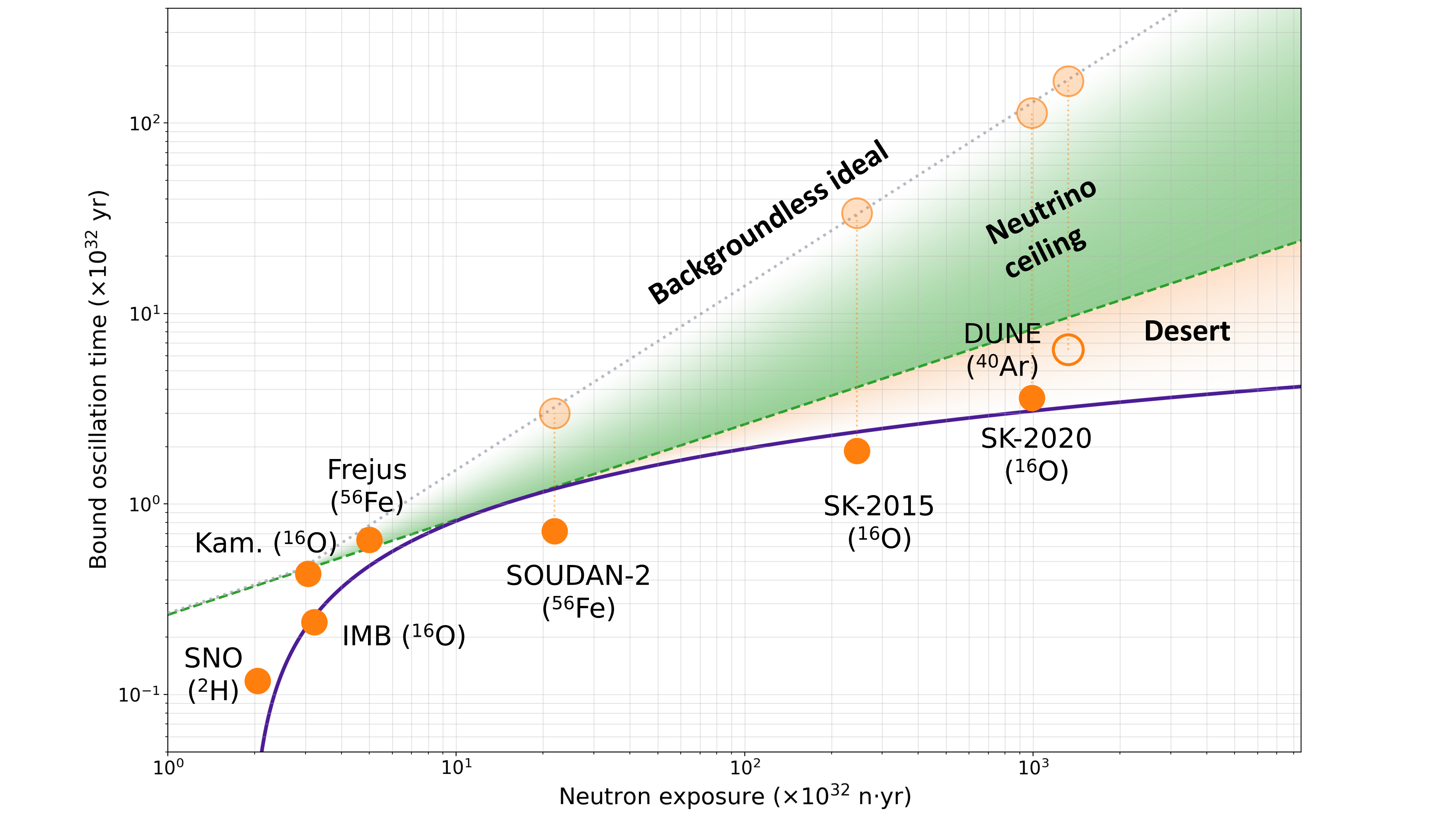}
    \\[2pt]
    \includegraphics[width=1.\columnwidth]{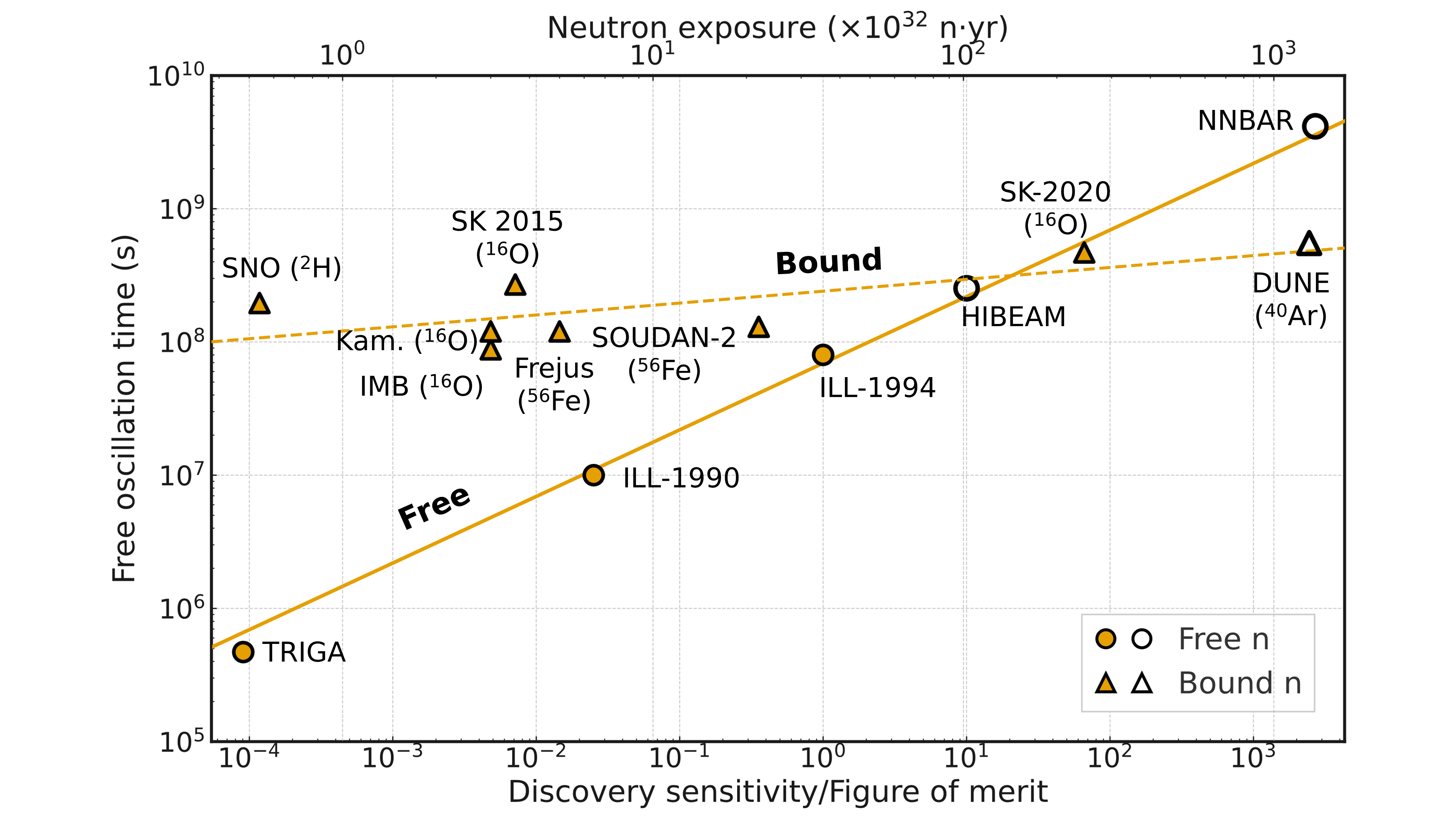}
    \caption{
    Top: Intranuclear $n\rightarrow\bar{n}$ experimental lower limits from SNO~\cite{SNO2017_bound}, Kamiokande (Kam.)~\cite{Kamiokande1986}, IMB~\cite{IMB1984_bound},Frejus~\cite{Frejus1990_bound}, SOUDAN-2~\cite{Soudan22002_bound}, and Super-Kamiokande (SK-2014~\cite{SuperK2015_bound}, SK-2020~\cite{Super-Kamiokande:2020bov}), are shown as solid orange points as a function of neutron exposure; a DUNE sensitivity from the horizontal drift technical design report~\cite{DUNE:2020ypp} is shown in an open circle. Open translucent circles are shown for hypothetical ideal sensitivities for large underground detectors assuming a single true event and $\sim 0$ background. Bottom: Free neutron oscillation time  as a function of the discovery sensitivity/figure-of-merit from past experiments at   TRIGA~\cite{Bressi1990_free} and the ILL (ILL-1990~\cite{BaldoCeolin1990_free} and ILL-1994~\cite{Baldo-Ceolin:1994hzw}). Projected sensitivities for the future experiments HIBEAM~\cite{Santoro:2023izd} and NNBAR~\cite{Santoro:2024lvc} are also given. Inferred free oscillation time from bound experiments are given for various exposures. Guide lines follow the free and bound data trends.}
  \label{fig:exp}
\end{figure}

\noindent Across the experimental landscape, matter stability searches, including bound $n\rightarrow \bar{n}$, have become a natural byproduct of large underground neutrino detectors (or, historically, \textit{vice-versa}). 

Figure~\ref{fig:exp}, top, shows experimental intranuclear $n\!\to\!\bar n$ lower exclusion limits versus neutron exposure. Data are given from Fréjus~\cite{Frejus1990_bound}, SOUDAN-2~\cite{Soudan22002_bound}, SNO~\cite{SNO2017_bound}, Super-K~\cite{SuperK2015_bound,Super-Kamiokande:2020bov}, Kamiokande~\cite{Kamiokande1986} and IMB~\cite{IMB1984}, together with a sensitivity projection from the future DUNE experiment~\cite{DUNE:2020ypp}\footnote{Published projected sensitivities from Hyper-Kamiokande~\cite{Hyper-Kamiokande:2018ofw} and JUNO~\cite{JUNO:2021vlw} have not yet been made.}. The relevant nuclei are stated. To guide the eye\footnote{This curve utilizes only experimentally derived exclusion limits. DUNE sensitivity limits are ignored.}, the solid purple curve approximately follows the previous limits, with a turnover in sensitivity highlighted for larger volume underground detectors. The dashed green line indicates the ideal statistics-limited trend $\propto \sqrt{\text{exposure}}$, and lies tangent to the solid purple curve, showing the empirical slope change at large exposures--this illustrates the discrepancy between an increase in mass and lowered exclusions due to proportionally lower signal efficiencies and higher neutrino backgrounds. The orange shaded region thus illustrates a hypothetical ``discovery desert'', where no discovery of $n\rightarrow\bar{n}$ can be convincingly made when dominated by background. Semi-transparent ``idealized'' points  are obtained with a TRolke-based\footnote{For each experiment we assume a single signal event is present with a $50\%$ signal efficiency \textit{a la} SNO and ILL, alongside a background of $\sim 0.01$ neutrino events. The signal efficiency and background rate errors are taken from the experimental values, if available. For each idealized limit, $1000$ pseudo-experiments are performed from Poissonian throws of the observed single event, and the median of the limit results from these pseudo-experiments is shown in each translucent point.} construction~\cite{Rolke:Lopez:Conrad:2005,Rolke:Lopez:2001,Lundberg:TRolke2:2010}--analogous to the statistical procedure used by Super-Kamiokande~\cite{SuperK2015_bound}--and then fitted to a gray dotted line. The green shaded region thus illustrates the gap between nominal background-limited performance and an idealized zero-background regime--a ``Neutrino Ceiling'' of sorts. Moving into this region experimentally can excite the possibility of discovery, but is still inevitably limited by at least~\textit{some} background. This, of course, is due to the fact that at sufficiently large exposures, atmospheric neutrino interactions can continue to be an irreducible background, imposing practical limits on any sensitivity. This is somewhat analogous to the ``neutrino floor’’ in WIMP searches~\cite{MonroeFisher2007,Billard2014,Ruppin2014}. This said, unlike the ``neutrino floor'' case, topological differentiation between the unique intranuclear $\bar{n}N$ annihilation signal and atmospheric backgrounds may be possible using novel and modern analysis methods and could hold the key to breaking this trend, including the use of improved reconstruction algorithms and perhaps direct applications of machine learning and computer vision.

For free–neutron searches operated in the quasi–free regime and at so-called zero background ($\ll 1$ event), as achieved at the last ILL experiment~\cite{Baldo-Ceolin:1994hzw}, discovery can be claimed with one event and the discovery sensitivity is determined by the figure-of-merit:
\begin{equation}
    F \;\equiv\; N\,\langle t^{2}\rangle.   
\end{equation}
\noindent Here, $N$ is the total number of neutrons accumulated during the run and $t$ is the neutron transit time in a magnetically shielded observation region.  In this limit, $P(n\!\to\!\bar n)\simeq (t/\tau_f)^2$. 

In contrast to bound neutron searches, due to intensity and beam infrastructure  requirements, free-beam searches have lain dormant, with the last search~\cite{Baldo-Ceolin:1994hzw} taking place in the 1990's. The European Spallation Source (ESS) opens a new discovery window for free $n\rightarrow \bar{n}$. The HIBEAM/NNBAR program is a multi-stage project to ultimately achieve a discovery sensitivity which is roughly three orders of magnitude beyond that achieved at the ILL~\cite{Addazi:2020nlz,Santoro:2023izd,Santoro:2024lvc}. 

Free oscillation time versus discovery sensitivity/figure-of-merit: $F\equiv N\langle t^{2}\rangle$ for free-beam searches (bottom axis) are shown in Figure~\ref{fig:exp}. The $F$ quantity is normalized such that $F=1$ for the last free search~\cite{Baldo-Ceolin:1994hzw}. Inferred free oscillation time versus neutron exposure (top axis) is shown for bound searches. Results from past free experiments at ILL and TRIGA~\cite{Bressi1990_free, BaldoCeolin1990_free, BaldoCeolin1990_free}\footnote{An earlier search at the ILL~\cite{Fidecaro1985_free} is not included as there was insufficient published information with which to calculate the figure-of-merit for that experiment.} as given are projected HIBEAM and NNBAR sensitivities~\cite{Santoro:2023izd, Santoro:2024lvc}.  The solid curve follows the zero-background scaling $\tau_f\propto\sqrt{F}$. The dashed guide line is an empirical power-law curve that approximately follows the bound points.

Figure~\ref{fig:exp}, bottom, shows the free (equivalent\footnote{Note that some past experiments utilized simplified methods of experimental lower limit extraction and associated free lifetime-equivalent conversion.}) transition time as a function of $F$ for previous free $n\!\to\!\bar n$ searches at the ILL~\cite{BaldoCeolin1990_free,Baldo-Ceolin:1994hzw} and Triga~\cite{Bressi1990_free}, together with the projected sensitivities for the two-stage HIBEAM/NNBAR program~\cite{Addazi:2020nlz,Santoro:2023izd,Santoro:2024lvc} the European Spallation Source. Up to three orders of magnitude in improvement in discovery sensitivity compared to the last search is possible. The largest challenge to HIBEAM/NNBAR performance is the achievement of the zero background condition at a spallation source.

\FloatBarrier 

\section*{Conversion of intranuclear results to free equivalents}
\noindent The so-called quasi-free regime for neutrons satisfies $\Delta E\, t \ll 1$ with $\Delta E \equiv E_n - E_{\bar n}$, where $E_n$ ($E_{\bar n}$) is the neutron (antineutron) energy and $t$ is the flight time. 
The two-level amplitude, $A^{f}(t)$, and probability, $P^{f}$,  are
\begin{align}
A^{f}(t) &= -i\, e^{-i \bar E t}\, \delta m\, t, & P^{f}(t) & = (\delta m\, t)^2,
\end{align}
where $\bar{E} \equiv (E_n + E_{\bar n})/2$. The microscopic mixing, $\delta m$, can be written in terms of Wilson coefficients at a scale $\mu$ and a hadronic matrix element for a set of operators $\mathcal{O}_i$ computed using lattice QCD~\cite{Rinaldi:2018osy,Rinaldi:2019thf} or more phenomenological approaches~\cite{Chodos:1974je,Chodos:1974pn,Chodos:1974dm}: $\delta m=\sum_i C_i(\mu)\,\langle \bar{n}\lvert \mathcal{O}_i(\mu)\rvert n\rangle$. The vacuum oscillation time is defined as $\tau_{f}=1/|\delta m|$.

For a neutron bound in a nucleus, taken schematically as a cavity in which the particles experience a constant potential, the effective transition amplitude $A^{b}(t)$ is given by: 
\begin{equation}
A^{b}(t)
= -\,i\,e^{-i\bar E t}\,
\delta m_{\rm eff}\,
\frac{1 - e^{-i\,\Delta E\, t - \Gamma_{\bar n} t/2}}{\Delta E - i\,\Gamma_{\bar n}/2}  \, ,
\end{equation}
and the probability  $P^{b}(t)=|A^{b}(t)|^2$, where $\Gamma_{\bar n}/2$ is the imaginary absorptive part of the optical potential. The mixing for the bound neutron case is defined as 

\begin{align}
\delta m_{\rm eff}
&= 
\sum_{i=1}^{7} C_i^{(9)}(\mu)\,
\langle \bar n|O_i^{(9)}(\mu)|n\rangle\,
F_i^{\rm nuc}
\!\left[1+c_2^{\rm had}\frac{\langle q^2\rangle_A}{\Lambda_\chi^2}+\cdots\right]
\label{eq:dm_eff_line1}\\[3pt]
&{}+ \frac{\langle q^2\rangle_A}{\Lambda_{11}^2}
\sum_{a} C_{a}^{(11)}(\mu)\,
\langle \bar n|O_{a}^{(11)}(\mu)|n\rangle\,
G_{a}^{\rm nuc}
+ \mathcal O\!\left(\frac{\langle q^2\rangle_A^{\,2}}{\Lambda^4}\right),
\label{eq:dm_eff_line2}
\end{align}

\noindent where $\langle q^2\rangle_A\!\sim\!p_F^2$ (set by Fermi motion), 
$c_2^{\rm had}$ parameterizes a hadronic (finite-size/nonlocal) slope of the $d{=}9$ matrix elements, and the explicit $q^2/\Lambda_{11}^2$ term captures derivative $d{=}11$ effects. The quantities $F_i^{\rm nuc}$ and $G_{a}^{\rm nuc}$ are nuclear overlap factors encoding many–body/nuclear–structure effects:
$F_i^{\mathrm{nuc}} \equiv \int d^3 r\, \psi_{\bar n}^{*}(\mathbf r)\,\mathcal F_i(\mathbf r)\,\psi_n(\mathbf r)$, 
where $\mathcal F_i(\mathbf{r})$ is a nuclear kernel; an analogous relation exists for $G_{a}^{\rm nuc}$.


The $R$ quantity, defined with respect to free and bound oscillation time, can also then be written:

\begin{equation}
R
= \frac{T_{b}}{\tau_{f}^2} 
=\frac{\Delta E^{2}+(\Gamma_{\bar n}/2)^{2}}{\Gamma_{\bar n}}\,
\frac{(\delta m)^2}{(\delta m_{\rm eff})^2}\,.
\end{equation}

\noindent In conventional analyses, $R$ is evaluated within a mean-field/optical-potential framework. The seven $d{=}9$ operators identified as providing $n\rightarrow \bar{n}$~\cite{Berezhiani:2018xsx} are taken as {coherent} with no momentum dependence of the mixing. It is assumed that in an EFT with a clear scale hierarchy (nucleon size $<$ inter-nucleon spacing), the short-distance $\Delta \mathcal{B}=2$ physics factorizes with the universal quadratic nucleon operator $n^TCn$ dominating with medium-dependent effects suppressed. This is hereafter referred to as the standard or baseline picture. 

Within the standard picture of $R$, determinations of $R$ carry a model dependence from optical potentials, annihilation widths, and nuclear structure. Quantifying these effects requires consistent hadronic inputs and many-body methods; some assessments span up to around 50\%~\cite{Phillips:2014fgb}.

\section{Beyond the baseline: \\in-medium reweighting and multi-operator interference}

\noindent Different operators carry different chiral and color--spin structures (e.g., $Q_L Q_L Q_L Q_L d_R d_R$ vs.\ $u_R d_R u_R d_R d_R d_R$), which transform differently under chiral rotations. In vacuum these differences can be immaterial if they contribute coherently to the same neutron–antineutron bilinear. In nuclear matter, however, chiral symmetry is explicitly and spontaneously broken; pion exchange, tensor forces, and spin–isospin correlations couple differently to left- and right-handed quark bilinears and to distinct color contractions. As a result, the in-medium renormalization of operator classes generally differs, leading to operator-dependent reweighting of the matrix elements even if the same operators add coherently in vacuum. In our notation this appears as an operator dependence in $F_i^{\rm nuc}$ and in analogous factors for other operators~\cite{Rao:1982gt,Rao:1984npb,Phillips:2014fgb,Buchoff:2015qwa,Friedman:2008es}.

Second, empirical evidence for medium modifications of quark-level structure comes from the EMC effect, which shows $\mathcal{O}(10\text{--}20\%)$ differences in quark momentum distributions between bound and free nucleons~\cite{Thomas:2018kcx}. If such linear-response modifications are present at high $Q^2$, it is not \emph{a priori} justified to assume that highly nonlinear six-quark operators experience no medium reweighting.

Third, the nuclear environment exhibits short-range correlations (SRCs) that strongly distort the high-momentum tail of the nucleon momentum distribution. Electron-scattering measurements indicate that these tails are generated by short-range $NN$ dynamics rather than a mean-field picture, with $np$ pairs dominating over $pp$ and $nn$ pairs~\cite{Arrington:2012,annurev:/content/journals/10.1146/annurev-nucl-102020-022253}. Such SRCs modify the overlap of six-quark operators with the relevant nuclear wavefunction components, again implying operator-dependent $F_i^{\rm nuc}$.

Finally, nuclear matrix elements are intrinsically nonlocal: a six-quark operator can act on quarks residing in different, spatially separated nucleons. Although the mean inter-nucleon distance ($\sim$1--2\,fm) exceeds the hadronic scale, Pauli blocking, scalar mean fields, and nucleon overlap alter short-distance nucleon structure and thus the effective matching of quark operators onto hadronic/nuclear degrees of freedom.

\subsection*{Multi-operator interference}
\noindent Taken together, these effects described  above imply that (i) the relative in-medium weights of different $\Delta\mathcal{B}=2$ operators can in general differ from their vacuum values, and (ii) constructive or destructive {multi-operator interference} in nuclei can arise even when such interference is absent for free neutrons. In the one–body amplitude, coherent interference is generic among the momentum–independent dim–9 and dim–11 pieces and the derivative contributions (both UV dim–11 and hadronic/in–medium). Predicting cancellations or enhancements therefore requires knowledge of Wilson coefficients (magnitudes and phases) {and} operator-resolved in-medium matrix elements.

Much theoretical work remains to be done in this area. However, models can be identified for which interference would be expected and the sensitivity of  $R$ to interference within a given scenario estimated.

\subsubsection{Left-right symmetry and multiple operators with $\Delta(\mathcal{B}-\mathcal{L})=2$}
\noindent The first interesting gauge theory where $n\rightarrow\bar{n}$ transitions were predicted at observable rates was left-right symmetric theory with quark-lepton unification~\cite{Mohapatra:1979ia}. It is shown that this model gives four kinds of independent operators with the dimension $d=9$, in which case there will definitely be different relative values for their contributions to oscillation versus nuclear decay. For example, in this model the source of the $\Delta \mathcal{B}=2$ interaction is the four $SU(4)_C$ 10-plet Higgs field of type $\epsilon^{aceg}\epsilon^{bdfh}\vec\Delta_{R, ab}\cdot\vec\Delta_{R, cd} \vec\Delta_{R, ef}\cdot\vec\Delta_{R, gh}$ and $\epsilon^{aceg}\epsilon^{bdfh}\vec\Delta_{L, ab}\cdot\vec\Delta_{L, cd}\vec\Delta_{R, ef}\cdot\vec\Delta_{R, gh}$ where $a,...h$ are the SU(4)$_C$ indices and vector signs correspond to SU(2)$_{L,R}$ groups. When the $\Delta_{R, 44}$ field acquires a vev, $v_{BL}$, they give rise to cubic coupled color sextet product operators of type
$v_{BL}\epsilon^{ikm}\epsilon^{jln}\Delta_{R, ij}\Delta_{R, kl}\Delta_{R, mn}$ and $v_{BL}\epsilon^{ikm}\epsilon^{jln}\Delta_{L, ij}\Delta_{L, kl}\Delta_{R, mn}$ . Couplings of these scalars to quarks give rise to the following operators which violate $B-L$ by two units:
%
\begin{multline}
  \Bigl[\frac{\lambda_Rv_{BL}}{M^6_{\Delta_R}}u_Rd_R u_R d_R d_Rd_R\\+\frac{\lambda_Rv_{BL}}{M^6_{\Delta_R}}d_Rd_R u_R u_R d_Rd_R+
		\frac{\lambda_Lv_{BL}}{M^4_{\Delta_L}M^2_{\Delta_R}}u_Ld_L u_L d_L d_Rd_R \\ +\frac{\lambda_Lv_{BL}}{M^4_{\Delta_L}M^2_{\Delta_R}}u_Lu_L d_L d_L d_Rd_R\Bigr]\,T_S  
\end{multline}
where $T_S=\epsilon_{cae}\epsilon_{dbf}+\epsilon_{cbf}\epsilon_{dae}+ \epsilon_{cbe}\epsilon_{daf}+\epsilon_{caf}\epsilon_{dbe}$, as given in the next subsection.
As a result, there are two independent RRR operators, and two LLR, which appear together. The transition and decay operators will therefore, in general, give different contributions, leading to the kind of effects being discussed here.
\subsubsection{Neutron oscillations from $R$-parity violation in SUSY}

\noindent In supersymmetric models with $R$-parity violation (RPV)~\cite{Barbier:2004ez}, the superpotential term,  
\begin{equation}
W_{\rm RPV} \;\supset\; \tfrac{1}{2} \, \lambda''_{ijk} \, U^c_i D^c_j D^c_k
\end{equation}
violates baryon number by one unit ($\Delta B = 1$). Here, $U^c_i$ and $D^c_j$ denote the right-handed up- and down-type quark superfields of generation $i$ and $j$, respectively, and $\lambda''_{ijk}$ is a dimensionless complex coupling controlling the strength of the baryon-number-violating interaction. The factor $1/2$ accounts for the antisymmetry of the color contraction between the two down-type superfields. Two insertions of this vertex, connected by virtual gluino exchange, generate $\Delta B = 2$ six-quark operators in the low-energy effective theory \cite{Calibbi:2016ukt}. Since the couplings $\lambda''_{ijk}$ are independent parameters, different color and flavor contractions produce a generic linear combination of operators in the neutron--antineutron EFT basis. For example, for some choice of parameters, two operators that are induced after SUSY breaking are:

\begin{equation}
\begin{aligned}
{\cal O}_1 &= u^a_R C d^b_R\, u^c_R C d^d_R\, d^e_R C d^f_R\, T^s_{abcdef},\\
{\cal O}_2 &= d^a_R C u^b_R\, d^c_L C d^d_L\, u^e_R C d^f_R\, T^s_{abcdef}.
\end{aligned}
\end{equation}
where $C$ is the charge conjugation matrix and $T^s_{abcdef} =\epsilon_{cae}\epsilon_{dbf}+\epsilon_{cbf}\epsilon_{dae}+ \epsilon_{cbe}\epsilon_{daf}+\epsilon_{caf}\epsilon_{dbe}$~\cite{Rao:1982plb}. Both terms arise from down squark exchange~\cite{Calibbi:2016ukt}. Prior to gauge symmetry breaking, these operators involve two different down flavors but once flavor mixings are included via SUSY breaking, the "flavor diagonal" $n\rightarrow\bar{n}$ operator emerges.
 
Thus, operator interference of the type studied here can arise in RPV SUSY. In the EFT basis, the effective dimension-9 operator coefficients are related to the squark and gluino masses  and SUSY breaking parameters $\Delta_{LR}$ and $\delta_{RR}$ as follows:
\begin{equation}
C_{\rm RPV} \;\sim\; \frac{\lambda''^2 \delta_{SUSY}}{m_{\tilde q}^4m_{\lambda_g}}.
\end{equation}
where $\tilde{q}$ and $\lambda_g$ respectively denote the squark and gluino masses and $\delta_{SUSY}$ involves the SUSY breaking parameters (see ~\cite{Calibbi:2016ukt} for detailed expressions for the SUSY breaking parameters).

\section{Monte Carlo Analysis of Operator Interference in RPV SUSY}
In order to quantify in-medium effects on neutron--antineutron oscillations, we perform a Monte Carlo study within a benchmark R-parity-violating supersymmetric framework. The analysis incorporates both nuclear dressing and the complex phases of the Wilson coefficients. This approach allows us to examine how these effects modify the in-medium single-operator transition rate and to identify parameter regions that lead to constructive or destructive interference.

The Wilson coefficients from the RPV SUSY scaling are taken as $C_1 = \frac{\lambda^{'' }_{113}\delta_{RR}\lambda^{'' }_{112}\delta_{RR}}{m_{\tilde q_d}^4 \, m_{\tilde g}}$ and $C_2 = \frac{(\lambda''_{113}\delta_{LR})^2}{m_{\tilde q_d}^4 \, m_{\tilde g}} \, e^{i (\phi_{C_2}-\phi_{C_1})}$, where $\phi_i$ is the complex phase of the $i$-th coefficient and $\delta$'s are SUSY breaking contributions. In this study, $C_1$ is taken real ($\phi_1 = 0$), and $C_2$ is allowed to vary in phase relative to $C_1$: $\phi_{C_2} - \phi_{C_1} \in [0, 2\pi]$. The vacuum matrix elements are fixed at $M_{f_i} = [100.0, 100.0]~\mathrm{MeV}$ for $i = 1,2$, and the in-medium matrix elements include nuclear dressing via $M_{a_i} = F_i^{\rm nuc} \, M_{f_i}$. The RPV SUSY benchmark parameters are $ m_{\tilde q_d} = 6~\mathrm{TeV},\; m_{\tilde g} = 9~\mathrm{TeV},\;$$\lambda''_{113}\delta_{RR} = 0.6$, $\lambda''_{112}\delta_{RR} = 0.4$ and $\lambda''_{113}\delta_{LR} = 0.6$. This makes it explicit that $C_1$ is a cross-term involving two different couplings, while $C_2$ is a single-coupling squared term.

With the benchmark choice of superpartner masses and couplings, the vacuum oscillation lifetime is $\tau_f = 1/|\delta m|$, with $\delta m = \sum_i C_i M_{f_i}$ 
For the baseline values 
this gives $\tau_f \simeq 1.92 \times 10^{11}~\mathrm{s}$. More generally, the oscillation time depends on the magnitude and relative phase of the Wilson coefficients as $|\delta m| = |C_1 M_{f_1} + C_2 M_{f_2} e^{i (\phi_{C_2}-\phi_{C_1})}|$.

To explore the effects of nuclear dressing and operator interference, a Monte Carlo scan of $N = 10^7$ random parameter points was performed, varying $|F_1^{\rm nuc}| \in [1.05, 1.30]$, $|F_2^{\rm nuc}| \in [1.10, 1.60]$, and $\phi_{C_2}-\phi_{C_1} \in [0, 2\pi]$. For each sampled point, the normalized rate $R/R_{\rm std}$ is computed relative to the single-operator reference.

\begin{figure}[htb]
\centering
\includegraphics[width=0.48\textwidth]{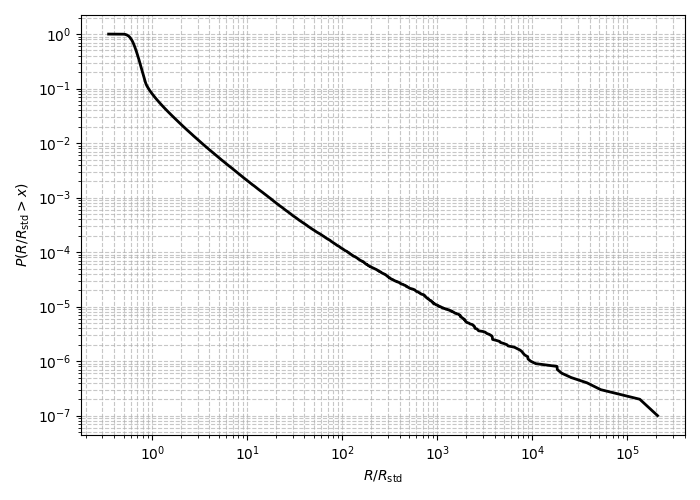}
\caption{Survival probability $P(R/R_{\rm std} > x)$ as a function of $R/R_{\rm std}$ computed from $10^7$ Monte Carlo samples.}
\label{fig:survival}
\end{figure}

From the ensemble of Monte Carlo samples, we computed the survival probability $P(R/R_{\rm std} > x) = 1 - \mathrm{CDF}(R/R_{\rm std})$, where CDF denotes the cumulative distribution function of the normalized rate. This function denotes the fraction of Monte Carlo samples yielding a normalized rate greater than $x$, i.e., the volume fraction of parameter space where the in-medium enhancement exceeds $x$ times the standard case, as shown in Figure~\ref{fig:survival}. Most samples cluster around moderate values, typically within one order of magnitude of the standard rate, while a small fraction of points form a long tail reaching up to $\mathcal{O}(10^3)$, with only a handful of points extending beyond $10^4$. Although the fraction of points leading to strong suppression is significantly smaller, such regions of parameter space are not negligible given that the relative complex phases of the Wilson coefficients are a priori unknown. Even small phase variations can therefore induce cancellations that reduce the effective in-medium oscillation rate by several orders of magnitude.

\begin{figure}[htb]
\centering
\includegraphics[width=0.48\textwidth]{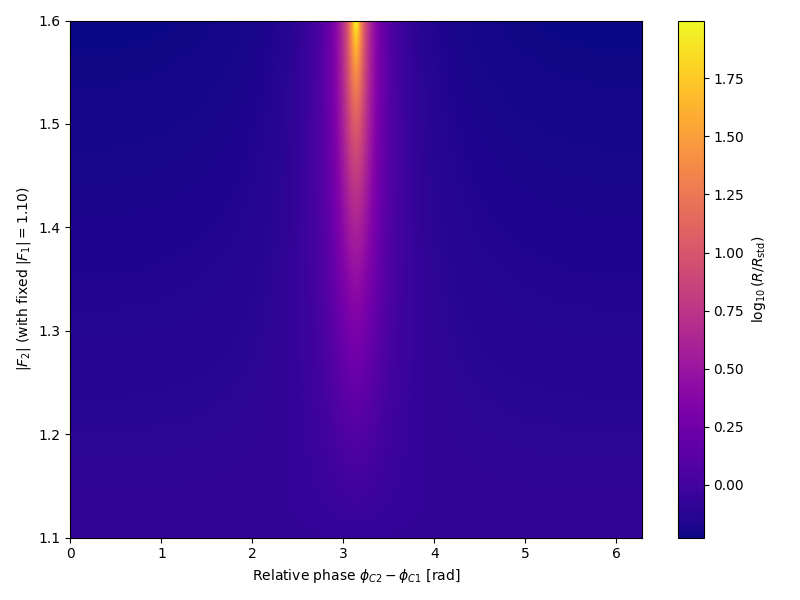}
\caption{Heatmap of $\log_{10}(R/R_{\rm std})$ as a function of the relative Wilson coefficient phase $\phi_{C_2}-\phi_{C_1}$ and $|F_2|$, for fixed $|F_1|=1.1$.}
\label{fig:heatmap_phase_F2}
\end{figure}

Figure~\ref{fig:heatmap_phase_F2} shows a two-dimensional heatmap of $R/R_{\rm std}$ as a function of the relative phase $\phi_{C_2}-\phi_{C_1} \in [0, 2\pi]$ and $|F_2|$ for fixed $|F_1|$, highlighting regions of constructive and destructive interference. The scan demonstrates that interference and nuclear dressing can enhance or suppress the $n$–$\bar{n}$ transition rate by several orders of magnitude, with constructive interference occurring for small relative phases, while amplitude-level destructive interference aligns near $\phi_{C_2}-\phi_{C_1} \approx \pi$ and is amplified for large $|F_2^{\rm nuc}|$. The full dynamic range of this effect extends beyond what is visible in the color scale of the heatmap, as extreme enhancements correspond to a very small fraction of the sampled parameter space.

\section*{Non-locality in $n\rightarrow \bar{n}$}

Non-locality can occur in a two-step $n\rightarrow \bar{n}$ transition in hidden sector scenarios~\cite{Berezhiani:2020vbe, Dvali:2024kk}. In Ref.~\cite{Berezhiani:2020vbe}, a neutron converts into a sterile neutron $n'$ which subsequently converts into an antineutron~\cite{Berezhiani:2020vbe}. 

Following Ref.~\cite{Berezhiani:2020vbe}, a two-step process, $n\!\to n'\!\to\!\bar n$, is considered in which the three levels are {degenerate in vacuum}, $E_n=E_{n'}=E_{\bar n}$. This approach can be applied for a range of magnetic field intervals in order to match equivalent fields in a hidden sector. Here, the case $B<10$~nT is considered. This corresponds to the field values needed to achieve the quasi-free condition in a traditional free $n\rightarrow \bar{n}$ search with beam neutrons~\cite{Davis:2016uyk}.   

For quasi-free neutrons, time--dependent perturbation theory gives the probability for short times:
\begin{equation}
\mathcal P_f^{(2)}(t)=\frac{\big|\varepsilon_{n n'}\varepsilon_{n'\bar n}\big|^{2}}{4}\,t^{4}.
\label{eq:t4-scaling}
\end{equation}
In this case, the free--beam probability grows as $t^{4}$, not $t^{2}$ as in the baseline one--step case. The appropriate free figure-of-merit and number of $\bar{n}$ annihilations at a beam experiment are, respectively, 
\begin{equation}
F^{(2)} \;\equiv\; \frac{N\,\langle t^{4}\rangle}{4} 
\label{eq:F4-def}
\end{equation}
and 
\begin{equation}
    Y_{\rm f}^{(2)} = \big|\varepsilon_{n n'}\varepsilon_{n'\bar n}\big|^{2}\,F^{(2)}_{\rm f}.
\end{equation}

Here, $N$ is the accumulated number of free neutrons in the run and $t$ is the neutron flight time determined by the spectrometer geometry~\cite{BaldoCeolin1990_free,Baldo-Ceolin:1994hzw}.

Inside nuclei the sterile level is detuned by the neutron mean field, while the $\bar n$ experiences a complex optical potential. In the optical--potential formalism~\cite{Dover:1982wv,Friedman:2008ef,Phillips:2014fgb}
the two--step intranuclear transition rate per bound neutron is
\begin{equation}
T_b^{(2)}\;=\;\frac{\Gamma\,\big|\varepsilon_{n n'}\varepsilon_{n'\bar n}\big|^{2}}
{\Delta_{n'}^{2}\,\big[\Delta^{2}+(\Gamma/2)^{2}\big]}\,,
\label{eq:Tb-2step}
\end{equation}
where $\Delta=(E_n-E_{\bar n})$ is the energy splitting of $n$ and $\bar n$ in the medium, $\Gamma\!\equiv\!-2\,\mathrm{Im}\,E_{\bar n}$ is the absorptive width of $\bar n$ in the medium, and $\Delta_{n'}\!\equiv\!E_n-E_{n'}$ is the in-medium non-degeneracy of the sterile neutron $n'$ with  ($\Delta_{n'}\!\equiv\!E_n-E_{n'} \sim \mathcal O(5\!-\!60)\,\mathrm{MeV}$ depending on peripherality/surface weighting~\cite{Friedman:2008ef}. For a detector with a total number of bound neutrons $N_b$ and live time $T$, the expected intranuclear yield is
\begin{equation}
    Y_{\rm b}^{(2)}=\big|\varepsilon_{n n'}\varepsilon_{n'\bar n}\big|^{2}\;
\frac{\Gamma}{\Delta_{n'}^{2}\,[\Delta^{2}+(\Gamma/2)^{2}]}\;
\big(N_b T).
\end{equation}

A representative one-year run at the ILL corresponds to  $N\sim 2\times 10^{18}$ and $\langle t^{4}\rangle\sim 10^{-4}\,\mathrm{s}^{4}$~\cite{Baldo-Ceolin:1994hzw}. Taking Super--Kamiokande as an example of a bound experiment, $(N_b T)_{\rm b}\sim (6\times 10^{33})\times (1\,\mathrm{yr}) \sim 1.9\times 10^{41}\,\mathrm{neutron\!-\!s}$ for 22.5~kt fiducial water over one year~\cite{Super-Kamiokande:2020bov}. For optical-model inputs, the following values are set: $\Delta\sim 60~\mathrm{MeV}$ and $\Gamma\sim 200~\mathrm{MeV}$~\cite{Dover:1982wv,Friedman:2008ef,Phillips:2014fgb}, with a surface-weighted sterile detuning of $\Delta_{n'}=5$~MeV. The ratio of expected yields can be estimated:

\begin{equation}
\frac{Y_{\rm f}^{(2)}}{Y_{\rm b}^{(2)}}\;=\;
\frac{F^{(2)}_{\rm f}}{(N_b T)}\;
\frac{\Delta_{n'}^{2}\,\big[\Delta^{2}+(\Gamma/2)^{2}\big]}{\Gamma}\ \sim\;1.8\times 10^{39} .
\end{equation}

This illustrates how bound neutron experiments are blind to this type of $n\rightarrow \bar{n}$ transition.

\section*{Conclusions}
The mapping of bound-neutron limits to free-neutron constraints for $n\rightarrow \bar{n}$ is traditionally made with a nucleus-specific conversion factor, estimated with a mean field theory approach that is assumed to be valid for all $n\rightarrow \bar{n}$ scenarios. However, multi-operator interference and non-locality can lead to variations in $R$ of orders of magnitude with respect to the standard approach. Depending on the theory scenario, either free or bound neutron searches can deliver higher sensitivity. This argues for a broad program in which high-sensitivity searches are made using both methods. Furthermore, in analogy to comparable systems (e.g. EDM, neutrinoless double beta decay), there is a need for a phenomenology program that integrates all available $\Delta B=2$ information (free and bound $n\rightarrow \bar{n}$, and dinucleon-decay searches) into a single, operator-based global analysis.

\bibliographystyle{apsrev4-2}

\section{Acknowledgements}
We gratefully acknowledge helpful discussions with Mike Snow, Arkady Vainshtein and Jean-Marc Richard. JLB would also like to thank Daisy Kalra, Linyan Wan, and Tyler Stokes for TRolke starting codes and associated discussion over many years. 

\bibliography{pap-nnbar_refs_fixed}

\begin{thebibliography}{92}%
\makeatletter
\providecommand \@ifxundefined [1]{%
 \@ifx{#1\undefined}
}%
\providecommand \@ifnum [1]{%
 \ifnum #1\expandafter \@firstoftwo
 \else \expandafter \@secondoftwo
 \fi
}%
\providecommand \@ifx [1]{%
 \ifx #1\expandafter \@firstoftwo
 \else \expandafter \@secondoftwo
 \fi
}%
\providecommand \natexlab [1]{#1}%
\providecommand \enquote  [1]{``#1''}%
\providecommand \bibnamefont  [1]{#1}%
\providecommand \bibfnamefont [1]{#1}%
\providecommand \citenamefont [1]{#1}%
\providecommand \href@noop [0]{\@secondoftwo}%
\providecommand \href [0]{\begingroup \@sanitize@url \@href}%
\providecommand \@href[1]{\@@startlink{#1}\@@href}%
\providecommand \@@href[1]{\endgroup#1\@@endlink}%
\providecommand \@sanitize@url [0]{\catcode `\\12\catcode `\$12\catcode `\&12\catcode `\#12\catcode `\^12\catcode `\_12\catcode `\%12\relax}%
\providecommand \@@startlink[1]{}%
\providecommand \@@endlink[0]{}%
\providecommand \url  [0]{\begingroup\@sanitize@url \@url }%
\providecommand \@url [1]{\endgroup\@href {#1}{\urlprefix }}%
\providecommand \urlprefix  [0]{URL }%
\providecommand \Eprint [0]{\href }%
\providecommand \doibase [0]{https://doi.org/}%
\providecommand \selectlanguage [0]{\@gobble}%
\providecommand \bibinfo  [0]{\@secondoftwo}%
\providecommand \bibfield  [0]{\@secondoftwo}%
\providecommand \translation [1]{[#1]}%
\providecommand \BibitemOpen [0]{}%
\providecommand \bibitemStop [0]{}%
\providecommand \bibitemNoStop [0]{.\EOS\space}%
\providecommand \EOS [0]{\spacefactor3000\relax}%
\providecommand \BibitemShut  [1]{\csname bibitem#1\endcsname}%
\let\auto@bib@innerbib\@empty
\bibitem [{\citenamefont {Riotto}\ and\ \citenamefont {Trodden}(1999)}]{Riotto:1999yt}%
  \BibitemOpen
  \bibfield  {author} {\bibinfo {author} {\bibfnamefont {A.}~\bibnamefont {Riotto}}\ and\ \bibinfo {author} {\bibfnamefont {M.}~\bibnamefont {Trodden}},\ }\href {https://doi.org/10.1146/annurev.nucl.49.1.35} {\bibfield  {journal} {\bibinfo  {journal} {Ann. Rev. Nucl. Part. Sci.}\ }\textbf {\bibinfo {volume} {49}},\ \bibinfo {pages} {35} (\bibinfo {year} {1999})},\ \Eprint {https://arxiv.org/abs/hep-ph/9901362} {arXiv:hep-ph/9901362} \BibitemShut {NoStop}%
\bibitem [{\citenamefont {Dine}\ and\ \citenamefont {Kusenko}(2003)}]{Dine:2003ax}%
  \BibitemOpen
  \bibfield  {author} {\bibinfo {author} {\bibfnamefont {M.}~\bibnamefont {Dine}}\ and\ \bibinfo {author} {\bibfnamefont {A.}~\bibnamefont {Kusenko}},\ }\href {https://doi.org/10.1103/RevModPhys.76.1} {\bibfield  {journal} {\bibinfo  {journal} {Rev. Mod. Phys.}\ }\textbf {\bibinfo {volume} {76}},\ \bibinfo {pages} {1} (\bibinfo {year} {2003})},\ \Eprint {https://arxiv.org/abs/hep-ph/0303065} {arXiv:hep-ph/0303065} \BibitemShut {NoStop}%
\bibitem [{\citenamefont {Babu}\ \emph {et~al.}(2006)\citenamefont {Babu}, \citenamefont {Mohapatra},\ and\ \citenamefont {Nasri}}]{Babu:2006psb}%
  \BibitemOpen
  \bibfield  {author} {\bibinfo {author} {\bibfnamefont {K.}~\bibnamefont {Babu}}, \bibinfo {author} {\bibfnamefont {R.}~\bibnamefont {Mohapatra}},\ and\ \bibinfo {author} {\bibfnamefont {S.}~\bibnamefont {Nasri}},\ }\href {https://doi.org/10.1103/PhysRevLett.97.131301} {\bibfield  {journal} {\bibinfo  {journal} {Phys. Rev. Lett.}\ }\textbf {\bibinfo {volume} {97}},\ \bibinfo {pages} {131301} (\bibinfo {year} {2006})}\BibitemShut {NoStop}%
\bibitem [{\citenamefont {Canetti}\ \emph {et~al.}(2012)\citenamefont {Canetti}, \citenamefont {Drewes},\ and\ \citenamefont {Shaposhnikov}}]{Canetti:2012zc}%
  \BibitemOpen
  \bibfield  {author} {\bibinfo {author} {\bibfnamefont {L.}~\bibnamefont {Canetti}}, \bibinfo {author} {\bibfnamefont {M.}~\bibnamefont {Drewes}},\ and\ \bibinfo {author} {\bibfnamefont {M.}~\bibnamefont {Shaposhnikov}},\ }\href {https://doi.org/10.1088/1367-2630/14/9/095012} {\bibfield  {journal} {\bibinfo  {journal} {New J. Phys.}\ }\textbf {\bibinfo {volume} {14}},\ \bibinfo {pages} {095012} (\bibinfo {year} {2012})},\ \Eprint {https://arxiv.org/abs/1204.4186} {arXiv:1204.4186} \BibitemShut {NoStop}%
\bibitem [{\citenamefont {Babu}\ \emph {et~al.}(2013)\citenamefont {Babu}, \citenamefont {Bhupal~Dev}, \citenamefont {Fortes},\ and\ \citenamefont {Mohapatra}}]{Babu:2013yca}%
  \BibitemOpen
  \bibfield  {author} {\bibinfo {author} {\bibfnamefont {K.~S.}\ \bibnamefont {Babu}}, \bibinfo {author} {\bibfnamefont {P.~S.}\ \bibnamefont {Bhupal~Dev}}, \bibinfo {author} {\bibfnamefont {E.~C. F.~S.}\ \bibnamefont {Fortes}},\ and\ \bibinfo {author} {\bibfnamefont {R.~N.}\ \bibnamefont {Mohapatra}},\ }\href {https://doi.org/10.1103/PhysRevD.87.115019} {\bibfield  {journal} {\bibinfo  {journal} {Phys. Rev. D}\ }\textbf {\bibinfo {volume} {87}},\ \bibinfo {pages} {115019} (\bibinfo {year} {2013})},\ \Eprint {https://arxiv.org/abs/1303.6918} {arXiv:1303.6918 [hep-ph]} \BibitemShut {NoStop}%
\bibitem [{\citenamefont {Sakharov}(1967)}]{Sakharov:1967dj}%
  \BibitemOpen
  \bibfield  {author} {\bibinfo {author} {\bibfnamefont {A.~D.}\ \bibnamefont {Sakharov}},\ }\href {https://doi.org/10.1070/PU1991v034n05ABEH002497} {\bibfield  {journal} {\bibinfo  {journal} {Pisma Zh. Eksp. Teor. Fiz.}\ }\textbf {\bibinfo {volume} {5}},\ \bibinfo {pages} {32} (\bibinfo {year} {1967})}\BibitemShut {NoStop}%
\bibitem [{\citenamefont {Weinberg}(1979)}]{Weinberg:1979sa}%
  \BibitemOpen
  \bibfield  {author} {\bibinfo {author} {\bibfnamefont {S.}~\bibnamefont {Weinberg}},\ }\href {https://doi.org/10.1103/PhysRevLett.43.1566} {\bibfield  {journal} {\bibinfo  {journal} {Phys. Rev. Lett.}\ }\textbf {\bibinfo {volume} {43}},\ \bibinfo {pages} {1566} (\bibinfo {year} {1979})}\BibitemShut {NoStop}%
\bibitem [{\citenamefont {Wilczek}\ and\ \citenamefont {Zee}(1979)}]{Wilczek:1979hc}%
  \BibitemOpen
  \bibfield  {author} {\bibinfo {author} {\bibfnamefont {F.}~\bibnamefont {Wilczek}}\ and\ \bibinfo {author} {\bibfnamefont {A.}~\bibnamefont {Zee}},\ }\href {https://doi.org/10.1103/PhysRevLett.43.1571} {\bibfield  {journal} {\bibinfo  {journal} {Phys. Rev. Lett.}\ }\textbf {\bibinfo {volume} {43}},\ \bibinfo {pages} {1571} (\bibinfo {year} {1979})}\BibitemShut {NoStop}%
\bibitem [{\citenamefont {'t~Hooft}(1976)}]{tHooft:1976rip}%
  \BibitemOpen
  \bibfield  {author} {\bibinfo {author} {\bibfnamefont {G.}~\bibnamefont {'t~Hooft}},\ }\href {https://doi.org/10.1103/PhysRevLett.37.8} {\bibfield  {journal} {\bibinfo  {journal} {Phys. Rev. Lett.}\ }\textbf {\bibinfo {volume} {37}},\ \bibinfo {pages} {8} (\bibinfo {year} {1976})}\BibitemShut {NoStop}%
\bibitem [{\citenamefont {Kuzmin}\ \emph {et~al.}(1985)\citenamefont {Kuzmin}, \citenamefont {Rubakov},\ and\ \citenamefont {Shaposhnikov}}]{Kuzmin:1985mm}%
  \BibitemOpen
  \bibfield  {author} {\bibinfo {author} {\bibfnamefont {V.~A.}\ \bibnamefont {Kuzmin}}, \bibinfo {author} {\bibfnamefont {V.~A.}\ \bibnamefont {Rubakov}},\ and\ \bibinfo {author} {\bibfnamefont {M.~E.}\ \bibnamefont {Shaposhnikov}},\ }\href {https://doi.org/10.1016/0370-2693(85)91028-7} {\bibfield  {journal} {\bibinfo  {journal} {Phys. Lett. B}\ }\textbf {\bibinfo {volume} {155}},\ \bibinfo {pages} {36} (\bibinfo {year} {1985})}\BibitemShut {NoStop}%
\bibitem [{\citenamefont {Abe}\ \emph {et~al.}(2017)\citenamefont {Abe} \emph {et~al.}}]{Abe:2017pde_pi0}%
  \BibitemOpen
  \bibfield  {author} {\bibinfo {author} {\bibfnamefont {K.}~\bibnamefont {Abe}} \emph {et~al.} (\bibinfo {collaboration} {Super-Kamiokande Collaboration}),\ }\href {https://doi.org/10.1103/PhysRevD.95.012004} {\bibfield  {journal} {\bibinfo  {journal} {Phys. Rev. D}\ }\textbf {\bibinfo {volume} {95}},\ \bibinfo {pages} {012004} (\bibinfo {year} {2017})},\ \Eprint {https://arxiv.org/abs/1610.03597} {arXiv:1610.03597 [hep-ex]} \BibitemShut {NoStop}%
\bibitem [{\citenamefont {Abe}\ \emph {et~al.}(2014)\citenamefont {Abe} \emph {et~al.}}]{Abe:2014nuK}%
  \BibitemOpen
  \bibfield  {author} {\bibinfo {author} {\bibfnamefont {K.}~\bibnamefont {Abe}} \emph {et~al.} (\bibinfo {collaboration} {Super-Kamiokande Collaboration}),\ }\href {https://doi.org/10.1103/PhysRevD.90.072005} {\bibfield  {journal} {\bibinfo  {journal} {Phys. Rev. D}\ }\textbf {\bibinfo {volume} {90}},\ \bibinfo {pages} {072005} (\bibinfo {year} {2014})},\ \Eprint {https://arxiv.org/abs/1408.1195} {arXiv:1408.1195 [hep-ex]} \BibitemShut {NoStop}%
\bibitem [{\citenamefont {Hirata}\ \emph {et~al.}(1989)\citenamefont {Hirata} \emph {et~al.}}]{Hirata:1989leptonMeson}%
  \BibitemOpen
  \bibfield  {author} {\bibinfo {author} {\bibfnamefont {K.~S.}\ \bibnamefont {Hirata}} \emph {et~al.} (\bibinfo {collaboration} {Kamiokande Collaboration}),\ }\href {https://doi.org/10.1016/0370-2693(89)90058-0} {\bibfield  {journal} {\bibinfo  {journal} {Phys. Lett. B}\ }\textbf {\bibinfo {volume} {220}},\ \bibinfo {pages} {308} (\bibinfo {year} {1989})}\BibitemShut {NoStop}%
\bibitem [{\citenamefont {Georgi}\ and\ \citenamefont {Glashow}(1974)}]{Georgi:1974gg}%
  \BibitemOpen
  \bibfield  {author} {\bibinfo {author} {\bibfnamefont {H.}~\bibnamefont {Georgi}}\ and\ \bibinfo {author} {\bibfnamefont {S.~L.}\ \bibnamefont {Glashow}},\ }\href {https://doi.org/10.1103/PhysRevLett.32.438} {\bibfield  {journal} {\bibinfo  {journal} {Phys. Rev. Lett.}\ }\textbf {\bibinfo {volume} {32}},\ \bibinfo {pages} {438} (\bibinfo {year} {1974})}\BibitemShut {NoStop}%
\bibitem [{\citenamefont {Langacker}(1981)}]{Langacker:1981GUTreview}%
  \BibitemOpen
  \bibfield  {author} {\bibinfo {author} {\bibfnamefont {P.}~\bibnamefont {Langacker}},\ }\href {https://doi.org/10.1016/0370-1573(81)90059-4} {\bibfield  {journal} {\bibinfo  {journal} {Phys. Rept.}\ }\textbf {\bibinfo {volume} {72}},\ \bibinfo {pages} {185} (\bibinfo {year} {1981})}\BibitemShut {NoStop}%
\bibitem [{\citenamefont {Hebecker}\ \emph {et~al.}(2024)\citenamefont {Hebecker}, \citenamefont {Hisano},\ and\ \citenamefont {Nagata}}]{Hebecker:2024PDG_GUTs}%
  \BibitemOpen
  \bibfield  {author} {\bibinfo {author} {\bibfnamefont {A.}~\bibnamefont {Hebecker}}, \bibinfo {author} {\bibfnamefont {J.}~\bibnamefont {Hisano}},\ and\ \bibinfo {author} {\bibfnamefont {N.}~\bibnamefont {Nagata}},\ }\href {https://doi.org/10.1103/PhysRevD.110.030001} {\bibinfo {title} {Grand unified theories}},\ \bibinfo {howpublished} {In \emph{Review of Particle Physics}, S. Navas \emph{et al.} (Particle Data Group)} (\bibinfo {year} {2024}),\ \bibinfo {note} {phys.\ Rev.\ D \textbf{110}, 030001 (2024). Revised August 2023.}\BibitemShut {Stop}%
\bibitem [{\citenamefont {Mohapatra}\ and\ \citenamefont {Marshak}(1980)}]{Mohapatra:1979ia}%
  \BibitemOpen
  \bibfield  {author} {\bibinfo {author} {\bibfnamefont {R.~N.}\ \bibnamefont {Mohapatra}}\ and\ \bibinfo {author} {\bibfnamefont {R.~E.}\ \bibnamefont {Marshak}},\ }\href {https://doi.org/10.1103/PhysRevLett.44.1316} {\bibfield  {journal} {\bibinfo  {journal} {Phys. Rev. Lett.}\ }\textbf {\bibinfo {volume} {44}},\ \bibinfo {pages} {1316} (\bibinfo {year} {1980})},\ \bibinfo {note} {[Erratum: Phys.Rev.Lett. 44, 1643 (1980)]}\BibitemShut {NoStop}%
\bibitem [{\citenamefont {Mohapatra}\ and\ \citenamefont {Senjanovic}(1981)}]{Mohapatra:1981prd}%
  \BibitemOpen
  \bibfield  {author} {\bibinfo {author} {\bibfnamefont {R.~N.}\ \bibnamefont {Mohapatra}}\ and\ \bibinfo {author} {\bibfnamefont {G.}~\bibnamefont {Senjanovic}},\ }\href {https://doi.org/10.1103/PhysRevD.23.165} {\bibfield  {journal} {\bibinfo  {journal} {Phys. Rev. D}\ }\textbf {\bibinfo {volume} {23}},\ \bibinfo {pages} {165} (\bibinfo {year} {1981})}\BibitemShut {NoStop}%
\bibitem [{\citenamefont {Nussinov}\ and\ \citenamefont {Shrock}(2002)}]{Nussinov:2002}%
  \BibitemOpen
  \bibfield  {author} {\bibinfo {author} {\bibfnamefont {S.}~\bibnamefont {Nussinov}}\ and\ \bibinfo {author} {\bibfnamefont {R.}~\bibnamefont {Shrock}},\ }\href {https://doi.org/10.1103/PhysRevLett.88.161601} {\bibfield  {journal} {\bibinfo  {journal} {Phys. Rev. Lett.}\ }\textbf {\bibinfo {volume} {88}},\ \bibinfo {pages} {161601} (\bibinfo {year} {2002})}\BibitemShut {NoStop}%
\bibitem [{\citenamefont {Barbier}\ \emph {et~al.}(2005)\citenamefont {Barbier} \emph {et~al.}}]{Barbier:2004ez}%
  \BibitemOpen
  \bibfield  {author} {\bibinfo {author} {\bibfnamefont {R.}~\bibnamefont {Barbier}} \emph {et~al.},\ }\href {https://doi.org/10.1016/j.physrep.2005.08.006} {\bibfield  {journal} {\bibinfo  {journal} {Phys. Rept.}\ }\textbf {\bibinfo {volume} {420}},\ \bibinfo {pages} {1} (\bibinfo {year} {2005})},\ \Eprint {https://arxiv.org/abs/hep-ph/0406039} {arXiv:hep-ph/0406039} \BibitemShut {NoStop}%
\bibitem [{\citenamefont {Phillips}\ \emph {et~al.}(2016)\citenamefont {Phillips}, \citenamefont {Snow}, \citenamefont {Babu}, \citenamefont {Banerjee}, \citenamefont {Baxter}, \citenamefont {Berezhiani},\ and\ \citenamefont {\textit{et al.}}}]{Phillips:2014fgb}%
  \BibitemOpen
  \bibfield  {author} {\bibinfo {author} {\bibfnamefont {D.~G.}\ \bibnamefont {Phillips}}, \bibinfo {author} {\bibfnamefont {W.~M.}\ \bibnamefont {Snow}}, \bibinfo {author} {\bibfnamefont {K.}~\bibnamefont {Babu}}, \bibinfo {author} {\bibfnamefont {S.}~\bibnamefont {Banerjee}}, \bibinfo {author} {\bibfnamefont {D.~V.}\ \bibnamefont {Baxter}}, \bibinfo {author} {\bibfnamefont {Z.}~\bibnamefont {Berezhiani}},\ and\ \bibinfo {author} {\bibnamefont {\textit{et al.}}},\ }\href {https://doi.org/10.1016/j.physrep.2015.11.001} {\bibfield  {journal} {\bibinfo  {journal} {Phys. Rept.}\ }\textbf {\bibinfo {volume} {612}},\ \bibinfo {pages} {1} (\bibinfo {year} {2016})},\ \Eprint {https://arxiv.org/abs/1410.1100} {arXiv:1410.1100} \BibitemShut {NoStop}%
\bibitem [{\citenamefont {Calibbi}\ \emph {et~al.}(2016)\citenamefont {Calibbi}, \citenamefont {Ferretti}, \citenamefont {Milstead}, \citenamefont {Petersson},\ and\ \citenamefont {P{\"o}ttgen}}]{Calibbi:2016ukt}%
  \BibitemOpen
  \bibfield  {author} {\bibinfo {author} {\bibfnamefont {L.}~\bibnamefont {Calibbi}}, \bibinfo {author} {\bibfnamefont {G.}~\bibnamefont {Ferretti}}, \bibinfo {author} {\bibfnamefont {D.}~\bibnamefont {Milstead}}, \bibinfo {author} {\bibfnamefont {C.}~\bibnamefont {Petersson}},\ and\ \bibinfo {author} {\bibfnamefont {R.}~\bibnamefont {P{\"o}ttgen}},\ }\href {https://doi.org/10.1007/JHEP05(2016)144} {\bibfield  {journal} {\bibinfo  {journal} {JHEP}\ }\textbf {\bibinfo {volume} {05}},\ \bibinfo {pages} {144}},\ \bibinfo {note} {[Erratum: JHEP 10, 195 (2017)]},\ \Eprint {https://arxiv.org/abs/1602.04821} {arXiv:1602.04821 [hep-ph]} \BibitemShut {NoStop}%
\bibitem [{\citenamefont {Fidecaro}\ \emph {et~al.}(1985)\citenamefont {Fidecaro}, \citenamefont {Fidecaro}, \citenamefont {Lanceri},\ and\ \citenamefont {et~al.}}]{Fidecaro1985_free}%
  \BibitemOpen
  \bibfield  {author} {\bibinfo {author} {\bibfnamefont {G.}~\bibnamefont {Fidecaro}}, \bibinfo {author} {\bibfnamefont {M.}~\bibnamefont {Fidecaro}}, \bibinfo {author} {\bibfnamefont {L.}~\bibnamefont {Lanceri}},\ and\ \bibinfo {author} {\bibnamefont {et~al.}},\ }\href {https://doi.org/10.1016/0370-2693(85)91367-X} {\bibfield  {journal} {\bibinfo  {journal} {Physics Letters B}\ }\textbf {\bibinfo {volume} {156}},\ \bibinfo {pages} {122} (\bibinfo {year} {1985})},\ \bibinfo {note} {first free-neutron search}\BibitemShut {NoStop}%
\bibitem [{\citenamefont {Bressi}\ \emph {et~al.}(1990)\citenamefont {Bressi}, \citenamefont {Calligarich}, \citenamefont {Cambiaghi},\ and\ \citenamefont {et~al.}}]{Bressi1990_free}%
  \BibitemOpen
  \bibfield  {author} {\bibinfo {author} {\bibfnamefont {G.}~\bibnamefont {Bressi}}, \bibinfo {author} {\bibfnamefont {E.}~\bibnamefont {Calligarich}}, \bibinfo {author} {\bibfnamefont {M.}~\bibnamefont {Cambiaghi}},\ and\ \bibinfo {author} {\bibnamefont {et~al.}},\ }\href {https://doi.org/10.1007/BF02789025} {\bibfield  {journal} {\bibinfo  {journal} {Il Nuovo Cimento A}\ }\textbf {\bibinfo {volume} {103}},\ \bibinfo {pages} {731} (\bibinfo {year} {1990})},\ \bibinfo {note} {tRIGA Mark II, Pavia}\BibitemShut {NoStop}%
\bibitem [{\citenamefont {Baldo-Ceolin}\ \emph {et~al.}(1994)\citenamefont {Baldo-Ceolin} \emph {et~al.}}]{Baldo-Ceolin:1994hzw}%
  \BibitemOpen
  \bibfield  {author} {\bibinfo {author} {\bibfnamefont {M.}~\bibnamefont {Baldo-Ceolin}} \emph {et~al.},\ }\href {https://doi.org/10.1007/BF01580321} {\bibfield  {journal} {\bibinfo  {journal} {Z. Phys. C}\ }\textbf {\bibinfo {volume} {63}},\ \bibinfo {pages} {409} (\bibinfo {year} {1994})}\BibitemShut {NoStop}%
\bibitem [{\citenamefont {Addazi}\ \emph {et~al.}(2021)\citenamefont {Addazi} \emph {et~al.}}]{Addazi:2020nlz}%
  \BibitemOpen
  \bibfield  {author} {\bibinfo {author} {\bibfnamefont {A.}~\bibnamefont {Addazi}} \emph {et~al.},\ }\href {https://doi.org/10.1088/1361-6471/abf429} {\bibfield  {journal} {\bibinfo  {journal} {J. Phys. G}\ }\textbf {\bibinfo {volume} {48}},\ \bibinfo {pages} {070501} (\bibinfo {year} {2021})},\ \Eprint {https://arxiv.org/abs/2006.04907} {arXiv:2006.04907 [physics.ins-det]} \BibitemShut {NoStop}%
\bibitem [{\citenamefont {Berger}\ \emph {et~al.}(1990)\citenamefont {Berger}, \citenamefont {Fr{\"o}hlich}, \citenamefont {M{\"o}nch},\ and\ \citenamefont {et~al. (Fr{\'e}jus~Collaboration)}}]{Frejus1990_bound}%
  \BibitemOpen
  \bibfield  {author} {\bibinfo {author} {\bibfnamefont {C.}~\bibnamefont {Berger}}, \bibinfo {author} {\bibfnamefont {M.}~\bibnamefont {Fr{\"o}hlich}}, \bibinfo {author} {\bibfnamefont {H.}~\bibnamefont {M{\"o}nch}},\ and\ \bibinfo {author} {\bibnamefont {et~al. (Fr{\'e}jus~Collaboration)}},\ }\href {https://doi.org/10.1016/0370-2693(90)90441-8} {\bibfield  {journal} {\bibinfo  {journal} {Physics Letters B}\ }\textbf {\bibinfo {volume} {240}},\ \bibinfo {pages} {237} (\bibinfo {year} {1990})}\BibitemShut {NoStop}%
\bibitem [{\citenamefont {Chung}\ \emph {et~al.}(2002)\citenamefont {Chung}, \citenamefont {Allison}, \citenamefont {Alner},\ and\ \citenamefont {et~al. (Soudan 2~Collaboration)}}]{Soudan22002_bound}%
  \BibitemOpen
  \bibfield  {author} {\bibinfo {author} {\bibfnamefont {J.}~\bibnamefont {Chung}}, \bibinfo {author} {\bibfnamefont {W.~W.~M.}\ \bibnamefont {Allison}}, \bibinfo {author} {\bibfnamefont {G.~J.}\ \bibnamefont {Alner}},\ and\ \bibinfo {author} {\bibnamefont {et~al. (Soudan 2~Collaboration)}},\ }\href {https://doi.org/10.1103/PhysRevD.66.032004} {\bibfield  {journal} {\bibinfo  {journal} {Physical Review D}\ }\textbf {\bibinfo {volume} {66}},\ \bibinfo {pages} {032004} (\bibinfo {year} {2002})}\BibitemShut {NoStop}%
\bibitem [{\citenamefont {Aharmim}\ and\ \citenamefont {et~al. (SNO~Collaboration)}(2017)}]{SNO2017_bound}%
  \BibitemOpen
  \bibfield  {author} {\bibinfo {author} {\bibfnamefont {B.}~\bibnamefont {Aharmim}}\ and\ \bibinfo {author} {\bibnamefont {et~al. (SNO~Collaboration)}},\ }\href {https://doi.org/10.1103/PhysRevD.96.092005} {\bibfield  {journal} {\bibinfo  {journal} {Physical Review D}\ }\textbf {\bibinfo {volume} {96}},\ \bibinfo {pages} {092005} (\bibinfo {year} {2017})},\ \bibinfo {note} {deuteron (D$_2$O) target}\BibitemShut {NoStop}%
\bibitem [{\citenamefont {Abe}\ and\ \citenamefont {et~al. (Super-Kamiokande~Collaboration)}(2015)}]{SuperK2015_bound}%
  \BibitemOpen
  \bibfield  {author} {\bibinfo {author} {\bibfnamefont {K.}~\bibnamefont {Abe}}\ and\ \bibinfo {author} {\bibnamefont {et~al. (Super-Kamiokande~Collaboration)}},\ }\href {https://doi.org/10.1103/PhysRevD.91.072006} {\bibfield  {journal} {\bibinfo  {journal} {Physical Review D}\ }\textbf {\bibinfo {volume} {91}},\ \bibinfo {pages} {072006} (\bibinfo {year} {2015})}\BibitemShut {NoStop}%
\bibitem [{\citenamefont {Abe}\ and\ \citenamefont {et~al. (Super-Kamiokande~Collaboration)}(2021)}]{SuperK2021_bound}%
  \BibitemOpen
  \bibfield  {author} {\bibinfo {author} {\bibfnamefont {K.}~\bibnamefont {Abe}}\ and\ \bibinfo {author} {\bibnamefont {et~al. (Super-Kamiokande~Collaboration)}},\ }\href {https://doi.org/10.1103/PhysRevD.103.012008} {\bibfield  {journal} {\bibinfo  {journal} {Physical Review D}\ }\textbf {\bibinfo {volume} {103}},\ \bibinfo {pages} {012008} (\bibinfo {year} {2021})}\BibitemShut {NoStop}%
\bibitem [{\citenamefont {Alberico}\ \emph {et~al.}(1982)\citenamefont {Alberico}, \citenamefont {Bottino},\ and\ \citenamefont {Molinari}}]{Alberico:1982nu}%
  \BibitemOpen
  \bibfield  {author} {\bibinfo {author} {\bibfnamefont {W.~M.}\ \bibnamefont {Alberico}}, \bibinfo {author} {\bibfnamefont {A.}~\bibnamefont {Bottino}},\ and\ \bibinfo {author} {\bibfnamefont {A.}~\bibnamefont {Molinari}},\ }\href {https://doi.org/10.1016/0370-2693(82)90493-2} {\bibfield  {journal} {\bibinfo  {journal} {Phys. Lett. B}\ }\textbf {\bibinfo {volume} {114}},\ \bibinfo {pages} {266} (\bibinfo {year} {1982})}\BibitemShut {NoStop}%
\bibitem [{\citenamefont {Dover}\ \emph {et~al.}(1983)\citenamefont {Dover}, \citenamefont {Gal},\ and\ \citenamefont {Richard}}]{Dover:1982wv}%
  \BibitemOpen
  \bibfield  {author} {\bibinfo {author} {\bibfnamefont {C.~B.}\ \bibnamefont {Dover}}, \bibinfo {author} {\bibfnamefont {A.}~\bibnamefont {Gal}},\ and\ \bibinfo {author} {\bibfnamefont {J.~M.}\ \bibnamefont {Richard}},\ }\href {https://doi.org/10.1103/PhysRevD.27.1090} {\bibfield  {journal} {\bibinfo  {journal} {Phys. Rev. D}\ }\textbf {\bibinfo {volume} {27}},\ \bibinfo {pages} {1090} (\bibinfo {year} {1983})}\BibitemShut {NoStop}%
\bibitem [{\citenamefont {Alberico}\ \emph {et~al.}(1984)\citenamefont {Alberico}, \citenamefont {Bernabeu}, \citenamefont {Bottino},\ and\ \citenamefont {Molinari}}]{Alberico:1984wk}%
  \BibitemOpen
  \bibfield  {author} {\bibinfo {author} {\bibfnamefont {W.~M.}\ \bibnamefont {Alberico}}, \bibinfo {author} {\bibfnamefont {J.}~\bibnamefont {Bernabeu}}, \bibinfo {author} {\bibfnamefont {A.}~\bibnamefont {Bottino}},\ and\ \bibinfo {author} {\bibfnamefont {A.}~\bibnamefont {Molinari}},\ }\href {https://doi.org/10.1016/0375-9474(84)90691-2} {\bibfield  {journal} {\bibinfo  {journal} {Nucl. Phys. A}\ }\textbf {\bibinfo {volume} {429}},\ \bibinfo {pages} {445} (\bibinfo {year} {1984})}\BibitemShut {NoStop}%
\bibitem [{\citenamefont {Dover}\ \emph {et~al.}(1985)\citenamefont {Dover}, \citenamefont {Gal},\ and\ \citenamefont {Richard}}]{Dover:1985hk}%
  \BibitemOpen
  \bibfield  {author} {\bibinfo {author} {\bibfnamefont {C.~B.}\ \bibnamefont {Dover}}, \bibinfo {author} {\bibfnamefont {A.}~\bibnamefont {Gal}},\ and\ \bibinfo {author} {\bibfnamefont {J.~M.}\ \bibnamefont {Richard}},\ }\href {https://doi.org/10.1103/PhysRevC.31.1423} {\bibfield  {journal} {\bibinfo  {journal} {Phys. Rev. C}\ }\textbf {\bibinfo {volume} {31}},\ \bibinfo {pages} {1423} (\bibinfo {year} {1985})}\BibitemShut {NoStop}%
\bibitem [{\citenamefont {Dover}\ \emph {et~al.}(1989)\citenamefont {Dover}, \citenamefont {Gal},\ and\ \citenamefont {Richard}}]{Dover:1989zz}%
  \BibitemOpen
  \bibfield  {author} {\bibinfo {author} {\bibfnamefont {C.~B.}\ \bibnamefont {Dover}}, \bibinfo {author} {\bibfnamefont {A.}~\bibnamefont {Gal}},\ and\ \bibinfo {author} {\bibfnamefont {J.~M.}\ \bibnamefont {Richard}},\ }\href {https://doi.org/10.1016/0168-9002(89)90239-8} {\bibfield  {journal} {\bibinfo  {journal} {Nucl. Instrum. Meth. A}\ }\textbf {\bibinfo {volume} {284}},\ \bibinfo {pages} {13} (\bibinfo {year} {1989})}\BibitemShut {NoStop}%
\bibitem [{\citenamefont {Alberico}\ \emph {et~al.}(1991)\citenamefont {Alberico}, \citenamefont {De~Pace},\ and\ \citenamefont {Pignone}}]{Alberico:1990ij}%
  \BibitemOpen
  \bibfield  {author} {\bibinfo {author} {\bibfnamefont {W.~M.}\ \bibnamefont {Alberico}}, \bibinfo {author} {\bibfnamefont {A.}~\bibnamefont {De~Pace}},\ and\ \bibinfo {author} {\bibfnamefont {M.}~\bibnamefont {Pignone}},\ }\href {https://doi.org/10.1016/0375-9474(91)90032-2} {\bibfield  {journal} {\bibinfo  {journal} {Nucl. Phys. A}\ }\textbf {\bibinfo {volume} {523}},\ \bibinfo {pages} {488} (\bibinfo {year} {1991})}\BibitemShut {NoStop}%
\bibitem [{\citenamefont {Hufner}\ and\ \citenamefont {Kopeliovich}(1998)}]{Hufner:1998gu}%
  \BibitemOpen
  \bibfield  {author} {\bibinfo {author} {\bibfnamefont {J.}~\bibnamefont {Hufner}}\ and\ \bibinfo {author} {\bibfnamefont {B.~Z.}\ \bibnamefont {Kopeliovich}},\ }\href {https://doi.org/10.1142/S0217732398002540} {\bibfield  {journal} {\bibinfo  {journal} {Mod. Phys. Lett. A}\ }\textbf {\bibinfo {volume} {13}},\ \bibinfo {pages} {2385} (\bibinfo {year} {1998})},\ \Eprint {https://arxiv.org/abs/hep-ph/9807210} {arXiv:hep-ph/9807210} \BibitemShut {NoStop}%
\bibitem [{\citenamefont {Friedman}\ and\ \citenamefont {Gal}(2008{\natexlab{a}})}]{Friedman:2008es}%
  \BibitemOpen
  \bibfield  {author} {\bibinfo {author} {\bibfnamefont {E.}~\bibnamefont {Friedman}}\ and\ \bibinfo {author} {\bibfnamefont {A.}~\bibnamefont {Gal}},\ }\href {https://doi.org/10.1103/PhysRevD.78.016002} {\bibfield  {journal} {\bibinfo  {journal} {Phys. Rev. D}\ }\textbf {\bibinfo {volume} {78}},\ \bibinfo {pages} {016002} (\bibinfo {year} {2008}{\natexlab{a}})},\ \Eprint {https://arxiv.org/abs/0803.3696} {arXiv:0803.3696 [hep-ph]} \BibitemShut {NoStop}%
\bibitem [{\citenamefont {Barrow}\ \emph {et~al.}(2022)\citenamefont {Barrow}, \citenamefont {Golubeva}, \citenamefont {Paryev},\ and\ \citenamefont {Richard}}]{Barrow:2022nxb}%
  \BibitemOpen
  \bibfield  {author} {\bibinfo {author} {\bibfnamefont {J.~L.}\ \bibnamefont {Barrow}}, \bibinfo {author} {\bibfnamefont {E.~S.}\ \bibnamefont {Golubeva}}, \bibinfo {author} {\bibfnamefont {E.}~\bibnamefont {Paryev}},\ and\ \bibinfo {author} {\bibfnamefont {J.-M.}\ \bibnamefont {Richard}},\ }\href {https://doi.org/10.1103/PhysRevC.105.065501} {\bibfield  {journal} {\bibinfo  {journal} {Phys. Rev. C}\ }\textbf {\bibinfo {volume} {105}},\ \bibinfo {pages} {065501} (\bibinfo {year} {2022})}\BibitemShut {NoStop}%
\bibitem [{\citenamefont {Barrow}\ \emph {et~al.}(2020)\citenamefont {Barrow}, \citenamefont {Golubeva}, \citenamefont {Paryev},\ and\ \citenamefont {Richard}}]{Barrow:2019viz}%
  \BibitemOpen
  \bibfield  {author} {\bibinfo {author} {\bibfnamefont {J.~L.}\ \bibnamefont {Barrow}}, \bibinfo {author} {\bibfnamefont {E.~S.}\ \bibnamefont {Golubeva}}, \bibinfo {author} {\bibfnamefont {E.}~\bibnamefont {Paryev}},\ and\ \bibinfo {author} {\bibfnamefont {J.-M.}\ \bibnamefont {Richard}},\ }\href {https://doi.org/10.1103/PhysRevD.101.036008} {\bibfield  {journal} {\bibinfo  {journal} {Phys. Rev. D}\ }\textbf {\bibinfo {volume} {101}},\ \bibinfo {pages} {036008} (\bibinfo {year} {2020})},\ \Eprint {https://arxiv.org/abs/1906.02833} {arXiv:1906.02833 [hep-ex]} \BibitemShut {NoStop}%
\bibitem [{\citenamefont {Pospelov}\ and\ \citenamefont {Ritz}(2005)}]{Pospelov:2005pr}%
  \BibitemOpen
  \bibfield  {author} {\bibinfo {author} {\bibfnamefont {M.}~\bibnamefont {Pospelov}}\ and\ \bibinfo {author} {\bibfnamefont {A.}~\bibnamefont {Ritz}},\ }\href {https://doi.org/10.1016/j.aop.2005.04.002} {\bibfield  {journal} {\bibinfo  {journal} {Annals of Physics}\ }\textbf {\bibinfo {volume} {318}},\ \bibinfo {pages} {119} (\bibinfo {year} {2005})}\BibitemShut {NoStop}%
\bibitem [{\citenamefont {Engel}\ \emph {et~al.}(2013)\citenamefont {Engel}, \citenamefont {Ramsey-Musolf},\ and\ \citenamefont {van Kolck}}]{Engel:2013lsa}%
  \BibitemOpen
  \bibfield  {author} {\bibinfo {author} {\bibfnamefont {J.}~\bibnamefont {Engel}}, \bibinfo {author} {\bibfnamefont {M.~J.}\ \bibnamefont {Ramsey-Musolf}},\ and\ \bibinfo {author} {\bibfnamefont {U.}~\bibnamefont {van Kolck}},\ }\href {https://doi.org/10.1016/j.ppnp.2013.03.003} {\bibfield  {journal} {\bibinfo  {journal} {Prog. Part. Nucl. Phys.}\ }\textbf {\bibinfo {volume} {71}},\ \bibinfo {pages} {21} (\bibinfo {year} {2013})}\BibitemShut {NoStop}%
\bibitem [{\citenamefont {Liu}\ \emph {et~al.}(2007)\citenamefont {Liu}, \citenamefont {Engel}, \citenamefont {Maekawa}, \citenamefont {Mukhopadhyay}, \citenamefont {Sahoo},\ and\ \citenamefont {Ramsey-Musolf}}]{LiuTimmermans:2007PRC}%
  \BibitemOpen
  \bibfield  {author} {\bibinfo {author} {\bibfnamefont {C.-P.}\ \bibnamefont {Liu}}, \bibinfo {author} {\bibfnamefont {J.}~\bibnamefont {Engel}}, \bibinfo {author} {\bibfnamefont {C.~M.}\ \bibnamefont {Maekawa}}, \bibinfo {author} {\bibfnamefont {T.~N.}\ \bibnamefont {Mukhopadhyay}}, \bibinfo {author} {\bibfnamefont {B.~K.}\ \bibnamefont {Sahoo}},\ and\ \bibinfo {author} {\bibfnamefont {M.~J.}\ \bibnamefont {Ramsey-Musolf}},\ }\href {https://doi.org/10.1103/PhysRevC.76.035503} {\bibfield  {journal} {\bibinfo  {journal} {Phys. Rev. C}\ }\textbf {\bibinfo {volume} {76}},\ \bibinfo {pages} {035503} (\bibinfo {year} {2007})}\BibitemShut {NoStop}%
\bibitem [{\citenamefont {Flambaum}\ and\ \citenamefont {Ginges}(2002)}]{FlambaumGinges:2002PRA}%
  \BibitemOpen
  \bibfield  {author} {\bibinfo {author} {\bibfnamefont {V.~V.}\ \bibnamefont {Flambaum}}\ and\ \bibinfo {author} {\bibfnamefont {J.~S.~M.}\ \bibnamefont {Ginges}},\ }\href {https://doi.org/10.1103/PhysRevA.65.032113} {\bibfield  {journal} {\bibinfo  {journal} {Phys. Rev. A}\ }\textbf {\bibinfo {volume} {65}},\ \bibinfo {pages} {032113} (\bibinfo {year} {2002})}\BibitemShut {NoStop}%
\bibitem [{\citenamefont {de~Vries}\ \emph {et~al.}(2011)\citenamefont {de~Vries}, \citenamefont {Higa}, \citenamefont {Liu}, \citenamefont {Mereghetti}, \citenamefont {Stetcu}, \citenamefont {Timmermans},\ and\ \citenamefont {van Kolck}}]{deVries:2011PRC}%
  \BibitemOpen
  \bibfield  {author} {\bibinfo {author} {\bibfnamefont {J.}~\bibnamefont {de~Vries}}, \bibinfo {author} {\bibfnamefont {R.}~\bibnamefont {Higa}}, \bibinfo {author} {\bibfnamefont {C.-P.}\ \bibnamefont {Liu}}, \bibinfo {author} {\bibfnamefont {E.}~\bibnamefont {Mereghetti}}, \bibinfo {author} {\bibfnamefont {I.}~\bibnamefont {Stetcu}}, \bibinfo {author} {\bibfnamefont {R.~G.~E.}\ \bibnamefont {Timmermans}},\ and\ \bibinfo {author} {\bibfnamefont {U.}~\bibnamefont {van Kolck}},\ }\href {https://doi.org/10.1103/PhysRevC.84.065501} {\bibfield  {journal} {\bibinfo  {journal} {Phys. Rev. C}\ }\textbf {\bibinfo {volume} {84}},\ \bibinfo {pages} {065501} (\bibinfo {year} {2011})},\ \Eprint {https://arxiv.org/abs/1109.3604} {arXiv:1109.3604} \BibitemShut {NoStop}%
\bibitem [{\citenamefont {Bsaisou}\ \emph {et~al.}(2015)\citenamefont {Bsaisou}, \citenamefont {de~Vries}, \citenamefont {Hanhart}, \citenamefont {Liebig}, \citenamefont {Meißner}, \citenamefont {Minossi}, \citenamefont {Nogga},\ and\ \citenamefont {Wirzba}}]{Bsaisou:2015AOP}%
  \BibitemOpen
  \bibfield  {author} {\bibinfo {author} {\bibfnamefont {J.}~\bibnamefont {Bsaisou}}, \bibinfo {author} {\bibfnamefont {J.}~\bibnamefont {de~Vries}}, \bibinfo {author} {\bibfnamefont {C.}~\bibnamefont {Hanhart}}, \bibinfo {author} {\bibfnamefont {S.}~\bibnamefont {Liebig}}, \bibinfo {author} {\bibfnamefont {U.-G.}\ \bibnamefont {Meißner}}, \bibinfo {author} {\bibfnamefont {D.}~\bibnamefont {Minossi}}, \bibinfo {author} {\bibfnamefont {A.}~\bibnamefont {Nogga}},\ and\ \bibinfo {author} {\bibfnamefont {A.}~\bibnamefont {Wirzba}},\ }\href {https://doi.org/10.1016/j.aop.2015.04.031} {\bibfield  {journal} {\bibinfo  {journal} {Annals of Physics}\ }\textbf {\bibinfo {volume} {359}},\ \bibinfo {pages} {317} (\bibinfo {year} {2015})},\ \Eprint {https://arxiv.org/abs/1411.5804} {arXiv:1411.5804} \BibitemShut {NoStop}%
\bibitem [{\citenamefont {Engel}\ and\ \citenamefont {Men{\'e}ndez}(2017)}]{Engel:2016xgb}%
  \BibitemOpen
  \bibfield  {author} {\bibinfo {author} {\bibfnamefont {J.}~\bibnamefont {Engel}}\ and\ \bibinfo {author} {\bibfnamefont {J.}~\bibnamefont {Men{\'e}ndez}},\ }\href {https://doi.org/10.1088/1361-6633/aa5bc5} {\bibfield  {journal} {\bibinfo  {journal} {Rep. Prog. Phys.}\ }\textbf {\bibinfo {volume} {80}},\ \bibinfo {pages} {046301} (\bibinfo {year} {2017})},\ \Eprint {https://arxiv.org/abs/1610.06548} {arXiv:1610.06548 [nucl-th]} \BibitemShut {NoStop}%
\bibitem [{\citenamefont {Cirigliano}\ \emph {et~al.}(2018)\citenamefont {Cirigliano}, \citenamefont {Dekens}, \citenamefont {de~Vries}, \citenamefont {Graesser},\ and\ \citenamefont {Mereghetti}}]{Cirigliano:2018Master}%
  \BibitemOpen
  \bibfield  {author} {\bibinfo {author} {\bibfnamefont {V.}~\bibnamefont {Cirigliano}}, \bibinfo {author} {\bibfnamefont {W.}~\bibnamefont {Dekens}}, \bibinfo {author} {\bibfnamefont {J.}~\bibnamefont {de~Vries}}, \bibinfo {author} {\bibfnamefont {M.~L.}\ \bibnamefont {Graesser}},\ and\ \bibinfo {author} {\bibfnamefont {E.}~\bibnamefont {Mereghetti}},\ }\href {https://doi.org/10.1007/JHEP12(2018)097} {\bibfield  {journal} {\bibinfo  {journal} {JHEP}\ }\textbf {\bibinfo {volume} {12}},\ \bibinfo {pages} {097}},\ \Eprint {https://arxiv.org/abs/1806.02780} {arXiv:1806.02780 [hep-ph]} \BibitemShut {NoStop}%
\bibitem [{\citenamefont {Fitzpatrick}\ \emph {et~al.}(2013)\citenamefont {Fitzpatrick}, \citenamefont {Haxton}, \citenamefont {Katz}, \citenamefont {Lubbers},\ and\ \citenamefont {Xu}}]{Fitzpatrick:2012ix}%
  \BibitemOpen
  \bibfield  {author} {\bibinfo {author} {\bibfnamefont {A.~L.}\ \bibnamefont {Fitzpatrick}}, \bibinfo {author} {\bibfnamefont {W.}~\bibnamefont {Haxton}}, \bibinfo {author} {\bibfnamefont {E.}~\bibnamefont {Katz}}, \bibinfo {author} {\bibfnamefont {N.}~\bibnamefont {Lubbers}},\ and\ \bibinfo {author} {\bibfnamefont {Y.}~\bibnamefont {Xu}},\ }\href {https://doi.org/10.1088/1475-7516/2013/02/004} {\bibfield  {journal} {\bibinfo  {journal} {JCAP}\ }\bibfield  {number} {\bibinfo  {number} { (02)},\ \bibinfo {pages} {004}},\ }\Eprint {https://arxiv.org/abs/1203.3542} {arXiv:1203.3542 [hep-ph]} \BibitemShut {NoStop}%
\bibitem [{\citenamefont {Anand}\ \emph {et~al.}(2014)\citenamefont {Anand}, \citenamefont {Fitzpatrick},\ and\ \citenamefont {Haxton}}]{Anand:2013yka}%
  \BibitemOpen
  \bibfield  {author} {\bibinfo {author} {\bibfnamefont {N.}~\bibnamefont {Anand}}, \bibinfo {author} {\bibfnamefont {A.~L.}\ \bibnamefont {Fitzpatrick}},\ and\ \bibinfo {author} {\bibfnamefont {W.~C.}\ \bibnamefont {Haxton}},\ }\href {https://doi.org/10.1103/PhysRevC.89.065501} {\bibfield  {journal} {\bibinfo  {journal} {Phys. Rev. C}\ }\textbf {\bibinfo {volume} {89}},\ \bibinfo {pages} {065501} (\bibinfo {year} {2014})},\ \Eprint {https://arxiv.org/abs/1308.6288} {arXiv:1308.6288 [hep-ph]} \BibitemShut {NoStop}%
\bibitem [{\citenamefont {Schneck}\ \emph {et~al.}(2015)\citenamefont {Schneck} \emph {et~al.}}]{Schneck:2015eqa}%
  \BibitemOpen
  \bibfield  {author} {\bibinfo {author} {\bibfnamefont {K.}~\bibnamefont {Schneck}} \emph {et~al.} (\bibinfo {collaboration} {SuperCDMS}),\ }\href {https://doi.org/10.1103/PhysRevD.91.092004} {\bibfield  {journal} {\bibinfo  {journal} {Phys. Rev. D}\ }\textbf {\bibinfo {volume} {91}},\ \bibinfo {pages} {092004} (\bibinfo {year} {2015})},\ \bibinfo {note} {includes interference among EFT operators},\ \Eprint {https://arxiv.org/abs/1503.03379} {arXiv:1503.03379 [astro-ph.CO]} \BibitemShut {NoStop}%
\bibitem [{\citenamefont {Calibbi}\ and\ \citenamefont {Signorelli}(2018)}]{Calibbi:2018RivNuovo}%
  \BibitemOpen
  \bibfield  {author} {\bibinfo {author} {\bibfnamefont {L.}~\bibnamefont {Calibbi}}\ and\ \bibinfo {author} {\bibfnamefont {G.}~\bibnamefont {Signorelli}},\ }\href {https://doi.org/10.1393/ncr/i2018-10144-0} {\bibfield  {journal} {\bibinfo  {journal} {Riv. Nuovo Cim.}\ }\textbf {\bibinfo {volume} {41}},\ \bibinfo {pages} {71} (\bibinfo {year} {2018})},\ \Eprint {https://arxiv.org/abs/1709.00294} {arXiv:1709.00294 [hep-ph]} \BibitemShut {NoStop}%
\bibitem [{\citenamefont {Kitano}\ \emph {et~al.}(2002)\citenamefont {Kitano}, \citenamefont {Koike},\ and\ \citenamefont {Okada}}]{Kitano:2002PRD}%
  \BibitemOpen
  \bibfield  {author} {\bibinfo {author} {\bibfnamefont {R.}~\bibnamefont {Kitano}}, \bibinfo {author} {\bibfnamefont {M.}~\bibnamefont {Koike}},\ and\ \bibinfo {author} {\bibfnamefont {Y.}~\bibnamefont {Okada}},\ }\href {https://doi.org/10.1103/PhysRevD.66.096002} {\bibfield  {journal} {\bibinfo  {journal} {Phys. Rev. D}\ }\textbf {\bibinfo {volume} {66}},\ \bibinfo {pages} {096002} (\bibinfo {year} {2002})},\ \Eprint {https://arxiv.org/abs/hep-ph/0203110} {arXiv:hep-ph/0203110 [hep-ph]} \BibitemShut {NoStop}%
\bibitem [{\citenamefont {Babu}\ and\ \citenamefont {Mohapatra}(2015)}]{Babu:2015axa}%
  \BibitemOpen
  \bibfield  {author} {\bibinfo {author} {\bibfnamefont {K.~S.}\ \bibnamefont {Babu}}\ and\ \bibinfo {author} {\bibfnamefont {R.~N.}\ \bibnamefont {Mohapatra}},\ }\href {https://doi.org/10.1103/PhysRevD.91.096009} {\bibfield  {journal} {\bibinfo  {journal} {Phys. Rev. D}\ }\textbf {\bibinfo {volume} {91}},\ \bibinfo {pages} {096009} (\bibinfo {year} {2015})},\ \Eprint {https://arxiv.org/abs/1504.01176} {arXiv:1504.01176} \BibitemShut {NoStop}%
\bibitem [{\citenamefont {Addazi}\ \emph {et~al.}(2017)\citenamefont {Addazi}, \citenamefont {Berezhiani},\ and\ \citenamefont {Kamyshkov}}]{Addazi:2016rgo}%
  \BibitemOpen
  \bibfield  {author} {\bibinfo {author} {\bibfnamefont {A.}~\bibnamefont {Addazi}}, \bibinfo {author} {\bibfnamefont {Z.}~\bibnamefont {Berezhiani}},\ and\ \bibinfo {author} {\bibfnamefont {Y.}~\bibnamefont {Kamyshkov}},\ }\href {https://doi.org/10.1140/epjc/s10052-017-4892-1} {\bibfield  {journal} {\bibinfo  {journal} {Eur. Phys. J. C}\ }\textbf {\bibinfo {volume} {77}},\ \bibinfo {pages} {301} (\bibinfo {year} {2017})},\ \Eprint {https://arxiv.org/abs/1607.00348} {arXiv:1607.00348} \BibitemShut {NoStop}%
\bibitem [{\citenamefont {Litos}\ \emph {et~al.}(2014)\citenamefont {Litos} \emph {et~al.}}]{Super-Kamiokande:2014hie}%
  \BibitemOpen
  \bibfield  {author} {\bibinfo {author} {\bibfnamefont {M.}~\bibnamefont {Litos}} \emph {et~al.} (\bibinfo {collaboration} {Super-Kamiokande}),\ }\href {https://doi.org/10.1103/PhysRevLett.112.131803} {\bibfield  {journal} {\bibinfo  {journal} {Phys. Rev. Lett.}\ }\textbf {\bibinfo {volume} {112}},\ \bibinfo {pages} {131803} (\bibinfo {year} {2014})}\BibitemShut {NoStop}%
\bibitem [{\citenamefont {Gustafson}\ \emph {et~al.}(2015)\citenamefont {Gustafson} \emph {et~al.}}]{Super-Kamiokande:2015jbb}%
  \BibitemOpen
  \bibfield  {author} {\bibinfo {author} {\bibfnamefont {J.}~\bibnamefont {Gustafson}} \emph {et~al.} (\bibinfo {collaboration} {Super-Kamiokande}),\ }\href {https://doi.org/10.1103/PhysRevD.91.072009} {\bibfield  {journal} {\bibinfo  {journal} {Phys. Rev. D}\ }\textbf {\bibinfo {volume} {91}},\ \bibinfo {pages} {072009} (\bibinfo {year} {2015})},\ \Eprint {https://arxiv.org/abs/1504.01041} {arXiv:1504.01041 [hep-ex]} \BibitemShut {NoStop}%
\bibitem [{\citenamefont {Takhistov}\ \emph {et~al.}(2015)\citenamefont {Takhistov} \emph {et~al.}}]{Super-Kamiokande:2015pys}%
  \BibitemOpen
  \bibfield  {author} {\bibinfo {author} {\bibfnamefont {V.}~\bibnamefont {Takhistov}} \emph {et~al.} (\bibinfo {collaboration} {Super-Kamiokande}),\ }\href {https://doi.org/10.1103/PhysRevLett.115.121803} {\bibfield  {journal} {\bibinfo  {journal} {Phys. Rev. Lett.}\ }\textbf {\bibinfo {volume} {115}},\ \bibinfo {pages} {121803} (\bibinfo {year} {2015})},\ \Eprint {https://arxiv.org/abs/1508.05530} {arXiv:1508.05530 [hep-ex]} \BibitemShut {NoStop}%
\bibitem [{\citenamefont {Takita}\ \emph {et~al.}(1986)\citenamefont {Takita}, \citenamefont {Arisaka}, \citenamefont {Kajita}, \citenamefont {Kifune}, \citenamefont {Koshiba}, \citenamefont {Miyano}, \citenamefont {Nakahata}, \citenamefont {Oyama}, \citenamefont {Sato}, \citenamefont {Suda}, \citenamefont {Suzuki}, \citenamefont {Takahashi},\ and\ \citenamefont {Totsuka}}]{Kamiokande1986}%
  \BibitemOpen
  \bibfield  {author} {\bibinfo {author} {\bibfnamefont {M.}~\bibnamefont {Takita}}, \bibinfo {author} {\bibfnamefont {K.}~\bibnamefont {Arisaka}}, \bibinfo {author} {\bibfnamefont {T.}~\bibnamefont {Kajita}}, \bibinfo {author} {\bibfnamefont {T.}~\bibnamefont {Kifune}}, \bibinfo {author} {\bibfnamefont {M.}~\bibnamefont {Koshiba}}, \bibinfo {author} {\bibfnamefont {K.}~\bibnamefont {Miyano}}, \bibinfo {author} {\bibfnamefont {M.}~\bibnamefont {Nakahata}}, \bibinfo {author} {\bibfnamefont {Y.}~\bibnamefont {Oyama}}, \bibinfo {author} {\bibfnamefont {N.}~\bibnamefont {Sato}}, \bibinfo {author} {\bibfnamefont {T.}~\bibnamefont {Suda}}, \bibinfo {author} {\bibfnamefont {A.}~\bibnamefont {Suzuki}}, \bibinfo {author} {\bibfnamefont {K.}~\bibnamefont {Takahashi}},\ and\ \bibinfo {author} {\bibfnamefont {Y.}~\bibnamefont {Totsuka}} (\bibinfo {collaboration} {Kamiokande}),\ }\href {https://doi.org/10.1103/PhysRevD.34.902} {\bibfield  {journal} {\bibinfo  {journal} {Phys. Rev. D}\ }\textbf {\bibinfo {volume} {34}},\
  \bibinfo {pages} {902} (\bibinfo {year} {1986})}\BibitemShut {NoStop}%
\bibitem [{\citenamefont {Jones}\ \emph {et~al.}(1984{\natexlab{a}})\citenamefont {Jones}, \citenamefont {Bionta}, \citenamefont {Blewitt},\ and\ \citenamefont {et~al. (IMB~Collaboration)}}]{IMB1984_bound}%
  \BibitemOpen
  \bibfield  {author} {\bibinfo {author} {\bibfnamefont {T.~W.}\ \bibnamefont {Jones}}, \bibinfo {author} {\bibfnamefont {R.~M.}\ \bibnamefont {Bionta}}, \bibinfo {author} {\bibfnamefont {G.}~\bibnamefont {Blewitt}},\ and\ \bibinfo {author} {\bibnamefont {et~al. (IMB~Collaboration)}},\ }\href {https://doi.org/10.1103/PhysRevLett.52.720} {\bibfield  {journal} {\bibinfo  {journal} {Physical Review Letters}\ }\textbf {\bibinfo {volume} {52}},\ \bibinfo {pages} {720} (\bibinfo {year} {1984}{\natexlab{a}})}\BibitemShut {NoStop}%
\bibitem [{\citenamefont {Abe}\ \emph {et~al.}(2021)\citenamefont {Abe} \emph {et~al.}}]{Super-Kamiokande:2020bov}%
  \BibitemOpen
  \bibfield  {author} {\bibinfo {author} {\bibfnamefont {K.}~\bibnamefont {Abe}} \emph {et~al.} (\bibinfo {collaboration} {Super-Kamiokande}),\ }\href {https://doi.org/10.1103/PhysRevD.103.012008} {\bibfield  {journal} {\bibinfo  {journal} {Phys. Rev. D}\ }\textbf {\bibinfo {volume} {103}},\ \bibinfo {pages} {012008} (\bibinfo {year} {2021})},\ \Eprint {https://arxiv.org/abs/2012.02607} {arXiv:2012.02607 [hep-ex]} \BibitemShut {NoStop}%
\bibitem [{\citenamefont {Abi}\ \emph {et~al.}(2020)\citenamefont {Abi} \emph {et~al.}}]{DUNE:2020ypp}%
  \BibitemOpen
  \bibfield  {author} {\bibinfo {author} {\bibfnamefont {B.}~\bibnamefont {Abi}} \emph {et~al.} (\bibinfo {collaboration} {DUNE}),\ }\href@noop {} {\emph {\bibinfo {title} {{Deep Underground Neutrino Experiment (DUNE), Far Detector Technical Design Report, Volume II: DUNE Physics}}}},\ \bibinfo {type} {Tech. Rep.}\ (\bibinfo {year} {2020})\ \Eprint {https://arxiv.org/abs/2002.03005} {arXiv:2002.03005 [hep-ex]} \BibitemShut {NoStop}%
\bibitem [{\citenamefont {Baldo{-}Ceolin}\ \emph {et~al.}(1990)\citenamefont {Baldo{-}Ceolin}, \citenamefont {Benetti}, \citenamefont {Bitter},\ and\ \citenamefont {et~al.}}]{BaldoCeolin1990_free}%
  \BibitemOpen
  \bibfield  {author} {\bibinfo {author} {\bibfnamefont {M.}~\bibnamefont {Baldo{-}Ceolin}}, \bibinfo {author} {\bibfnamefont {P.}~\bibnamefont {Benetti}}, \bibinfo {author} {\bibfnamefont {T.}~\bibnamefont {Bitter}},\ and\ \bibinfo {author} {\bibnamefont {et~al.}},\ }\href {https://doi.org/10.1016/0370-2693(90)90601-2} {\bibfield  {journal} {\bibinfo  {journal} {Physics Letters B}\ }\textbf {\bibinfo {volume} {236}},\ \bibinfo {pages} {95} (\bibinfo {year} {1990})},\ \bibinfo {note} {iLL (Grenoble) beam experiment}\BibitemShut {NoStop}%
\bibitem [{\citenamefont {Santoro}\ \emph {et~al.}(2025)\citenamefont {Santoro} \emph {et~al.}}]{Santoro:2023izd}%
  \BibitemOpen
  \bibfield  {author} {\bibinfo {author} {\bibfnamefont {V.}~\bibnamefont {Santoro}} \emph {et~al.},\ }\href {https://doi.org/10.1088/1361-6471/adc8c2} {\bibfield  {journal} {\bibinfo  {journal} {J. Phys. G}\ }\textbf {\bibinfo {volume} {52}},\ \bibinfo {pages} {040501} (\bibinfo {year} {2025})},\ \Eprint {https://arxiv.org/abs/2311.08326} {arXiv:2311.08326 [physics.ins-det]} \BibitemShut {NoStop}%
\bibitem [{\citenamefont {Santoro}\ \emph {et~al.}(2024)\citenamefont {Santoro} \emph {et~al.}}]{Santoro:2024lvc}%
  \BibitemOpen
  \bibfield  {author} {\bibinfo {author} {\bibfnamefont {V.}~\bibnamefont {Santoro}} \emph {et~al.},\ }\href {https://doi.org/10.3233/jnr-230951} {\bibfield  {journal} {\bibinfo  {journal} {J. Neutron Res.}\ }\textbf {\bibinfo {volume} {25}},\ \bibinfo {pages} {315} (\bibinfo {year} {2024})}\BibitemShut {NoStop}%
\bibitem [{\citenamefont {Jones}\ \emph {et~al.}(1984{\natexlab{b}})\citenamefont {Jones}, \citenamefont {Bionta}, \citenamefont {Blewitt}, \citenamefont {Bratton}, \citenamefont {Cortez}, \citenamefont {Errede}, \citenamefont {Foster}, \citenamefont {Gajewski}, \citenamefont {Ganezer}, \citenamefont {Goldhaber}, \citenamefont {Haines}, \citenamefont {Kielczewska}, \citenamefont {Kropp}, \citenamefont {Learned}, \citenamefont {Lehmann}, \citenamefont {LoSecco}, \citenamefont {Park}, \citenamefont {Reines}, \citenamefont {Schultz}, \citenamefont {Shumard}, \citenamefont {Sinclair}, \citenamefont {Sobel}, \citenamefont {Stone}, \citenamefont {Sulak}, \citenamefont {Svoboda}, \citenamefont {Velde},\ and\ \citenamefont {Wuest}}]{IMB1984}%
  \BibitemOpen
  \bibfield  {author} {\bibinfo {author} {\bibfnamefont {T.~W.}\ \bibnamefont {Jones}}, \bibinfo {author} {\bibfnamefont {R.~M.}\ \bibnamefont {Bionta}}, \bibinfo {author} {\bibfnamefont {G.}~\bibnamefont {Blewitt}}, \bibinfo {author} {\bibfnamefont {C.~B.}\ \bibnamefont {Bratton}}, \bibinfo {author} {\bibfnamefont {B.~G.}\ \bibnamefont {Cortez}}, \bibinfo {author} {\bibfnamefont {S.}~\bibnamefont {Errede}}, \bibinfo {author} {\bibfnamefont {G.~W.}\ \bibnamefont {Foster}}, \bibinfo {author} {\bibfnamefont {W.}~\bibnamefont {Gajewski}}, \bibinfo {author} {\bibfnamefont {K.~S.}\ \bibnamefont {Ganezer}}, \bibinfo {author} {\bibfnamefont {M.}~\bibnamefont {Goldhaber}}, \bibinfo {author} {\bibfnamefont {T.~J.}\ \bibnamefont {Haines}}, \bibinfo {author} {\bibfnamefont {D.}~\bibnamefont {Kielczewska}}, \bibinfo {author} {\bibfnamefont {W.~R.}\ \bibnamefont {Kropp}}, \bibinfo {author} {\bibfnamefont {J.~G.}\ \bibnamefont {Learned}}, \bibinfo {author} {\bibfnamefont {E.}~\bibnamefont {Lehmann}}, \bibinfo {author}
  {\bibfnamefont {J.~M.}\ \bibnamefont {LoSecco}}, \bibinfo {author} {\bibfnamefont {H.~S.}\ \bibnamefont {Park}}, \bibinfo {author} {\bibfnamefont {F.}~\bibnamefont {Reines}}, \bibinfo {author} {\bibfnamefont {J.}~\bibnamefont {Schultz}}, \bibinfo {author} {\bibfnamefont {E.}~\bibnamefont {Shumard}}, \bibinfo {author} {\bibfnamefont {D.}~\bibnamefont {Sinclair}}, \bibinfo {author} {\bibfnamefont {H.~W.}\ \bibnamefont {Sobel}}, \bibinfo {author} {\bibfnamefont {J.~L.}\ \bibnamefont {Stone}}, \bibinfo {author} {\bibfnamefont {L.~R.}\ \bibnamefont {Sulak}}, \bibinfo {author} {\bibfnamefont {R.}~\bibnamefont {Svoboda}}, \bibinfo {author} {\bibfnamefont {J.~C. V.~D.}\ \bibnamefont {Velde}},\ and\ \bibinfo {author} {\bibfnamefont {C.}~\bibnamefont {Wuest}} (\bibinfo {collaboration} {Irvine--Michigan--Brookhaven (IMB)}),\ }\href {https://doi.org/10.1103/PhysRevLett.52.720} {\bibfield  {journal} {\bibinfo  {journal} {Phys. Rev. Lett.}\ }\textbf {\bibinfo {volume} {52}},\ \bibinfo {pages} {720} (\bibinfo {year}
  {1984}{\natexlab{b}})}\BibitemShut {NoStop}%
\bibitem [{\citenamefont {Abe}\ \emph {et~al.}(2018)\citenamefont {Abe} \emph {et~al.}}]{Hyper-Kamiokande:2018ofw}%
  \BibitemOpen
  \bibfield  {author} {\bibinfo {author} {\bibfnamefont {K.}~\bibnamefont {Abe}} \emph {et~al.} (\bibinfo {collaboration} {Hyper-Kamiokande Proto-Collaboration}),\ }\href@noop {} {\emph {\bibinfo {title} {Hyper-Kamiokande Design Report}}},\ \bibinfo {type} {Tech. Rep.}\ (\bibinfo {year} {2018})\ \bibinfo {note} {v2, 28 Nov 2018},\ \Eprint {https://arxiv.org/abs/1805.04163} {arXiv:1805.04163 [physics.ins-det]} \BibitemShut {NoStop}%
\bibitem [{\citenamefont {Abusleme}\ \emph {et~al.}(2022)\citenamefont {Abusleme} \emph {et~al.}}]{JUNO:2021vlw}%
  \BibitemOpen
  \bibfield  {author} {\bibinfo {author} {\bibfnamefont {A.}~\bibnamefont {Abusleme}} \emph {et~al.} (\bibinfo {collaboration} {JUNO}),\ }\href {https://doi.org/10.1016/j.ppnp.2021.103927} {\bibfield  {journal} {\bibinfo  {journal} {Prog. Part. Nucl. Phys.}\ }\textbf {\bibinfo {volume} {123}},\ \bibinfo {pages} {103927} (\bibinfo {year} {2022})},\ \Eprint {https://arxiv.org/abs/2104.02565} {arXiv:2104.02565 [hep-ex]} \BibitemShut {NoStop}%
\bibitem [{\citenamefont {Rolke}\ \emph {et~al.}(2005)\citenamefont {Rolke}, \citenamefont {L{\'o}pez},\ and\ \citenamefont {Conrad}}]{Rolke:Lopez:Conrad:2005}%
  \BibitemOpen
  \bibfield  {author} {\bibinfo {author} {\bibfnamefont {W.~A.}\ \bibnamefont {Rolke}}, \bibinfo {author} {\bibfnamefont {A.~M.}\ \bibnamefont {L{\'o}pez}},\ and\ \bibinfo {author} {\bibfnamefont {J.}~\bibnamefont {Conrad}},\ }\href {https://doi.org/10.1016/j.nima.2005.05.068} {\bibfield  {journal} {\bibinfo  {journal} {Nuclear Instruments and Methods in Physics Research Section A}\ }\textbf {\bibinfo {volume} {551}},\ \bibinfo {pages} {493} (\bibinfo {year} {2005})},\ \Eprint {https://arxiv.org/abs/physics/0403059} {arXiv:physics/0403059} \BibitemShut {NoStop}%
\bibitem [{\citenamefont {Rolke}\ and\ \citenamefont {L{\'o}pez}(2001)}]{Rolke:Lopez:2001}%
  \BibitemOpen
  \bibfield  {author} {\bibinfo {author} {\bibfnamefont {W.~A.}\ \bibnamefont {Rolke}}\ and\ \bibinfo {author} {\bibfnamefont {A.~M.}\ \bibnamefont {L{\'o}pez}},\ }\href {https://doi.org/10.1016/S0168-9002(00)00935-9} {\bibfield  {journal} {\bibinfo  {journal} {Nuclear Instruments and Methods in Physics Research Section A}\ }\textbf {\bibinfo {volume} {458}},\ \bibinfo {pages} {745} (\bibinfo {year} {2001})},\ \Eprint {https://arxiv.org/abs/hep-ph/0005187} {arXiv:hep-ph/0005187} \BibitemShut {NoStop}%
\bibitem [{\citenamefont {Lundberg}\ \emph {et~al.}(2010)\citenamefont {Lundberg}, \citenamefont {Conrad}, \citenamefont {Rolke},\ and\ \citenamefont {L{\'o}pez}}]{Lundberg:TRolke2:2010}%
  \BibitemOpen
  \bibfield  {author} {\bibinfo {author} {\bibfnamefont {J.}~\bibnamefont {Lundberg}}, \bibinfo {author} {\bibfnamefont {J.}~\bibnamefont {Conrad}}, \bibinfo {author} {\bibfnamefont {W.}~\bibnamefont {Rolke}},\ and\ \bibinfo {author} {\bibfnamefont {A.}~\bibnamefont {L{\'o}pez}},\ }\href {https://doi.org/10.1016/j.cpc.2009.11.001} {\bibfield  {journal} {\bibinfo  {journal} {Computer Physics Communications}\ }\textbf {\bibinfo {volume} {181}},\ \bibinfo {pages} {683} (\bibinfo {year} {2010})},\ \Eprint {https://arxiv.org/abs/0907.3450} {arXiv:0907.3450 [physics.data-an]} \BibitemShut {NoStop}%
\bibitem [{\citenamefont {Monroe}\ and\ \citenamefont {Fisher}(2007)}]{MonroeFisher2007}%
  \BibitemOpen
  \bibfield  {author} {\bibinfo {author} {\bibfnamefont {J.}~\bibnamefont {Monroe}}\ and\ \bibinfo {author} {\bibfnamefont {P.}~\bibnamefont {Fisher}},\ }\href {https://doi.org/10.1103/PhysRevD.76.033007} {\bibfield  {journal} {\bibinfo  {journal} {Phys. Rev. D}\ }\textbf {\bibinfo {volume} {76}},\ \bibinfo {pages} {033007} (\bibinfo {year} {2007})},\ \Eprint {https://arxiv.org/abs/0706.3019} {arXiv:0706.3019} \BibitemShut {NoStop}%
\bibitem [{\citenamefont {Billard}\ \emph {et~al.}(2014)\citenamefont {Billard}, \citenamefont {Strigari},\ and\ \citenamefont {Figueroa-Feliciano}}]{Billard2014}%
  \BibitemOpen
  \bibfield  {author} {\bibinfo {author} {\bibfnamefont {J.}~\bibnamefont {Billard}}, \bibinfo {author} {\bibfnamefont {L.}~\bibnamefont {Strigari}},\ and\ \bibinfo {author} {\bibfnamefont {E.}~\bibnamefont {Figueroa-Feliciano}},\ }\href {https://doi.org/10.1103/PhysRevD.89.023524} {\bibfield  {journal} {\bibinfo  {journal} {Phys. Rev. D}\ }\textbf {\bibinfo {volume} {89}},\ \bibinfo {pages} {023524} (\bibinfo {year} {2014})},\ \Eprint {https://arxiv.org/abs/1307.5458} {arXiv:1307.5458} \BibitemShut {NoStop}%
\bibitem [{\citenamefont {Ruppin}\ \emph {et~al.}(2014)\citenamefont {Ruppin}, \citenamefont {Billard}, \citenamefont {Figueroa-Feliciano},\ and\ \citenamefont {Strigari}}]{Ruppin2014}%
  \BibitemOpen
  \bibfield  {author} {\bibinfo {author} {\bibfnamefont {F.}~\bibnamefont {Ruppin}}, \bibinfo {author} {\bibfnamefont {J.}~\bibnamefont {Billard}}, \bibinfo {author} {\bibfnamefont {E.}~\bibnamefont {Figueroa-Feliciano}},\ and\ \bibinfo {author} {\bibfnamefont {L.}~\bibnamefont {Strigari}},\ }\href {https://doi.org/10.1103/PhysRevD.90.083510} {\bibfield  {journal} {\bibinfo  {journal} {Phys. Rev. D}\ }\textbf {\bibinfo {volume} {90}},\ \bibinfo {pages} {083510} (\bibinfo {year} {2014})},\ \Eprint {https://arxiv.org/abs/1408.3581} {arXiv:1408.3581} \BibitemShut {NoStop}%
\bibitem [{\citenamefont {Rinaldi}\ \emph {et~al.}(2019{\natexlab{a}})\citenamefont {Rinaldi}, \citenamefont {Syritsyn}, \citenamefont {Wagman}, \citenamefont {Buchoff}, \citenamefont {Schroeder},\ and\ \citenamefont {Wasem}}]{Rinaldi:2018osy}%
  \BibitemOpen
  \bibfield  {author} {\bibinfo {author} {\bibfnamefont {E.}~\bibnamefont {Rinaldi}}, \bibinfo {author} {\bibfnamefont {S.}~\bibnamefont {Syritsyn}}, \bibinfo {author} {\bibfnamefont {M.~L.}\ \bibnamefont {Wagman}}, \bibinfo {author} {\bibfnamefont {M.~I.}\ \bibnamefont {Buchoff}}, \bibinfo {author} {\bibfnamefont {C.}~\bibnamefont {Schroeder}},\ and\ \bibinfo {author} {\bibfnamefont {J.}~\bibnamefont {Wasem}},\ }\href {https://doi.org/10.1103/PhysRevLett.122.162001} {\bibfield  {journal} {\bibinfo  {journal} {Phys. Rev. Lett.}\ }\textbf {\bibinfo {volume} {122}},\ \bibinfo {pages} {162001} (\bibinfo {year} {2019}{\natexlab{a}})},\ \Eprint {https://arxiv.org/abs/1809.00246} {arXiv:1809.00246 [hep-lat]} \BibitemShut {NoStop}%
\bibitem [{\citenamefont {Rinaldi}\ \emph {et~al.}(2019{\natexlab{b}})\citenamefont {Rinaldi}, \citenamefont {Syritsyn}, \citenamefont {Wagman}, \citenamefont {Buchoff}, \citenamefont {Schroeder},\ and\ \citenamefont {Wasem}}]{Rinaldi:2019thf}%
  \BibitemOpen
  \bibfield  {author} {\bibinfo {author} {\bibfnamefont {E.}~\bibnamefont {Rinaldi}}, \bibinfo {author} {\bibfnamefont {S.}~\bibnamefont {Syritsyn}}, \bibinfo {author} {\bibfnamefont {M.~L.}\ \bibnamefont {Wagman}}, \bibinfo {author} {\bibfnamefont {M.~I.}\ \bibnamefont {Buchoff}}, \bibinfo {author} {\bibfnamefont {C.}~\bibnamefont {Schroeder}},\ and\ \bibinfo {author} {\bibfnamefont {J.}~\bibnamefont {Wasem}},\ }\href {https://doi.org/10.1103/PhysRevD.99.074510} {\bibfield  {journal} {\bibinfo  {journal} {Phys. Rev. D}\ }\textbf {\bibinfo {volume} {99}},\ \bibinfo {pages} {074510} (\bibinfo {year} {2019}{\natexlab{b}})},\ \Eprint {https://arxiv.org/abs/1901.07519} {arXiv:1901.07519 [hep-lat]} \BibitemShut {NoStop}%
\bibitem [{\citenamefont {Chodos}\ \emph {et~al.}(1974{\natexlab{a}})\citenamefont {Chodos}, \citenamefont {Jaffe}, \citenamefont {Johnson}, \citenamefont {Thorn},\ and\ \citenamefont {Weisskopf}}]{Chodos:1974je}%
  \BibitemOpen
  \bibfield  {author} {\bibinfo {author} {\bibfnamefont {A.}~\bibnamefont {Chodos}}, \bibinfo {author} {\bibfnamefont {R.~L.}\ \bibnamefont {Jaffe}}, \bibinfo {author} {\bibfnamefont {K.}~\bibnamefont {Johnson}}, \bibinfo {author} {\bibfnamefont {C.~B.}\ \bibnamefont {Thorn}},\ and\ \bibinfo {author} {\bibfnamefont {V.~F.}\ \bibnamefont {Weisskopf}},\ }\href {https://doi.org/10.1103/PhysRevD.9.3471} {\bibfield  {journal} {\bibinfo  {journal} {Phys. Rev. D}\ }\textbf {\bibinfo {volume} {9}},\ \bibinfo {pages} {3471} (\bibinfo {year} {1974}{\natexlab{a}})}\BibitemShut {NoStop}%
\bibitem [{\citenamefont {Chodos}\ \emph {et~al.}(1974{\natexlab{b}})\citenamefont {Chodos}, \citenamefont {Jaffe}, \citenamefont {Johnson},\ and\ \citenamefont {Thorn}}]{Chodos:1974pn}%
  \BibitemOpen
  \bibfield  {author} {\bibinfo {author} {\bibfnamefont {A.}~\bibnamefont {Chodos}}, \bibinfo {author} {\bibfnamefont {R.~L.}\ \bibnamefont {Jaffe}}, \bibinfo {author} {\bibfnamefont {K.}~\bibnamefont {Johnson}},\ and\ \bibinfo {author} {\bibfnamefont {C.~B.}\ \bibnamefont {Thorn}},\ }\href {https://doi.org/10.1103/PhysRevD.10.2599} {\bibfield  {journal} {\bibinfo  {journal} {Phys. Rev. D}\ }\textbf {\bibinfo {volume} {10}},\ \bibinfo {pages} {2599} (\bibinfo {year} {1974}{\natexlab{b}})}\BibitemShut {NoStop}%
\bibitem [{\citenamefont {Chodos}\ and\ \citenamefont {Thorn}(1974)}]{Chodos:1974dm}%
  \BibitemOpen
  \bibfield  {author} {\bibinfo {author} {\bibfnamefont {A.}~\bibnamefont {Chodos}}\ and\ \bibinfo {author} {\bibfnamefont {C.~B.}\ \bibnamefont {Thorn}},\ }\href {https://doi.org/10.1016/0370-2693(74)90402-X} {\bibfield  {journal} {\bibinfo  {journal} {Phys. Lett. B}\ }\textbf {\bibinfo {volume} {53}},\ \bibinfo {pages} {359} (\bibinfo {year} {1974})}\BibitemShut {NoStop}%
\bibitem [{\citenamefont {Berezhiani}\ and\ \citenamefont {Vainshtein}(2019)}]{Berezhiani:2018xsx}%
  \BibitemOpen
  \bibfield  {author} {\bibinfo {author} {\bibfnamefont {Z.}~\bibnamefont {Berezhiani}}\ and\ \bibinfo {author} {\bibfnamefont {A.}~\bibnamefont {Vainshtein}},\ }\href {https://doi.org/10.1016/j.physletb.2018.11.014} {\bibfield  {journal} {\bibinfo  {journal} {Phys. Lett. B}\ }\textbf {\bibinfo {volume} {788}},\ \bibinfo {pages} {58} (\bibinfo {year} {2019})},\ \Eprint {https://arxiv.org/abs/1809.00997} {arXiv:1809.00997 [hep-ph]} \BibitemShut {NoStop}%
\bibitem [{\citenamefont {Rao}\ and\ \citenamefont {Shrock}(1982{\natexlab{a}})}]{Rao:1982gt}%
  \BibitemOpen
  \bibfield  {author} {\bibinfo {author} {\bibfnamefont {S.}~\bibnamefont {Rao}}\ and\ \bibinfo {author} {\bibfnamefont {R.~E.}\ \bibnamefont {Shrock}},\ }\href {https://doi.org/10.1016/0370-2693(82)90333-1} {\bibfield  {journal} {\bibinfo  {journal} {Phys. Lett. B}\ }\textbf {\bibinfo {volume} {116}},\ \bibinfo {pages} {238} (\bibinfo {year} {1982}{\natexlab{a}})}\BibitemShut {NoStop}%
\bibitem [{\citenamefont {Rao}\ and\ \citenamefont {Shrock}(1984)}]{Rao:1984npb}%
  \BibitemOpen
  \bibfield  {author} {\bibinfo {author} {\bibfnamefont {S.}~\bibnamefont {Rao}}\ and\ \bibinfo {author} {\bibfnamefont {R.~E.}\ \bibnamefont {Shrock}},\ }\href {https://doi.org/10.1016/0550-3213(84)90365-1} {\bibfield  {journal} {\bibinfo  {journal} {Nucl. Phys. B}\ }\textbf {\bibinfo {volume} {232}},\ \bibinfo {pages} {143} (\bibinfo {year} {1984})}\BibitemShut {NoStop}%
\bibitem [{\citenamefont {Buchoff}\ and\ \citenamefont {Wagman}(2016)}]{Buchoff:2015qwa}%
  \BibitemOpen
  \bibfield  {author} {\bibinfo {author} {\bibfnamefont {M.~I.}\ \bibnamefont {Buchoff}}\ and\ \bibinfo {author} {\bibfnamefont {M.~L.}\ \bibnamefont {Wagman}},\ }\href {https://doi.org/10.1103/PhysRevD.93.016005} {\bibfield  {journal} {\bibinfo  {journal} {Phys. Rev. D}\ }\textbf {\bibinfo {volume} {93}},\ \bibinfo {pages} {016005} (\bibinfo {year} {2016})},\ \Eprint {https://arxiv.org/abs/1506.00647} {arXiv:1506.00647} \BibitemShut {NoStop}%
\bibitem [{\citenamefont {Thomas}(2019)}]{Thomas:2018kcx}%
  \BibitemOpen
  \bibfield  {author} {\bibinfo {author} {\bibfnamefont {A.~W.}\ \bibnamefont {Thomas}},\ }\href {https://doi.org/10.1142/S0218301318400013} {\bibfield  {journal} {\bibinfo  {journal} {Int. J. Mod. Phys. E}\ }\textbf {\bibinfo {volume} {27}},\ \bibinfo {pages} {1840001} (\bibinfo {year} {2019})},\ \Eprint {https://arxiv.org/abs/1809.06622} {arXiv:1809.06622 [hep-ph]} \BibitemShut {NoStop}%
\bibitem [{\citenamefont {Arrington}\ \emph {et~al.}(2012)\citenamefont {Arrington}, \citenamefont {Higinbotham}, \citenamefont {Rosner},\ and\ \citenamefont {Weinstein}}]{Arrington:2012}%
  \BibitemOpen
  \bibfield  {author} {\bibinfo {author} {\bibfnamefont {J.}~\bibnamefont {Arrington}}, \bibinfo {author} {\bibfnamefont {D.~W.}\ \bibnamefont {Higinbotham}}, \bibinfo {author} {\bibfnamefont {G.}~\bibnamefont {Rosner}},\ and\ \bibinfo {author} {\bibfnamefont {M.}~\bibnamefont {Weinstein}},\ }\href {https://doi.org/10.1016/j.ppnp.2012.04.002} {\bibfield  {journal} {\bibinfo  {journal} {Prog. Part. Nucl. Phys.}\ }\textbf {\bibinfo {volume} {67}},\ \bibinfo {pages} {898} (\bibinfo {year} {2012})}\BibitemShut {NoStop}%
\bibitem [{\citenamefont {Arrington}\ \emph {et~al.}(2022)\citenamefont {Arrington}, \citenamefont {Fomin},\ and\ \citenamefont {Schmidt}}]{annurev:/content/journals/10.1146/annurev-nucl-102020-022253}%
  \BibitemOpen
  \bibfield  {author} {\bibinfo {author} {\bibfnamefont {J.}~\bibnamefont {Arrington}}, \bibinfo {author} {\bibfnamefont {N.}~\bibnamefont {Fomin}},\ and\ \bibinfo {author} {\bibfnamefont {A.}~\bibnamefont {Schmidt}},\ }\href {https://doi.org/https://doi.org/10.1146/annurev-nucl-102020-022253} {\bibfield  {journal} {\bibinfo  {journal} {Annual Review of Nuclear and Particle Science}\ }\textbf {\bibinfo {volume} {72}},\ \bibinfo {pages} {307} (\bibinfo {year} {2022})}\BibitemShut {NoStop}%
\bibitem [{\citenamefont {Rao}\ and\ \citenamefont {Shrock}(1982{\natexlab{b}})}]{Rao:1982plb}%
  \BibitemOpen
  \bibfield  {author} {\bibinfo {author} {\bibfnamefont {S.}~\bibnamefont {Rao}}\ and\ \bibinfo {author} {\bibfnamefont {R.~E.}\ \bibnamefont {Shrock}},\ }\href {https://doi.org/10.1016/0370-2693(82)90333-1} {\bibfield  {journal} {\bibinfo  {journal} {Phys. Lett. B}\ }\textbf {\bibinfo {volume} {116}},\ \bibinfo {pages} {238} (\bibinfo {year} {1982}{\natexlab{b}})}\BibitemShut {NoStop}%
\bibitem [{\citenamefont {Berezhiani}(2021)}]{Berezhiani:2020vbe}%
  \BibitemOpen
  \bibfield  {author} {\bibinfo {author} {\bibfnamefont {Z.}~\bibnamefont {Berezhiani}},\ }\href {https://doi.org/10.1140/epjc/s10052-020-08824-9} {\bibfield  {journal} {\bibinfo  {journal} {Eur. Phys. J. C}\ }\textbf {\bibinfo {volume} {81}},\ \bibinfo {pages} {33} (\bibinfo {year} {2021})},\ \Eprint {https://arxiv.org/abs/2002.05609} {arXiv:2002.05609 [hep-ph]} \BibitemShut {NoStop}%
\bibitem [{\citenamefont {Dvali}\ \emph {et~al.}(2024)\citenamefont {Dvali}, \citenamefont {Ettengruber},\ and\ \citenamefont {Stuhlfauth}}]{Dvali:2024kk}%
  \BibitemOpen
  \bibfield  {author} {\bibinfo {author} {\bibfnamefont {G.}~\bibnamefont {Dvali}}, \bibinfo {author} {\bibfnamefont {M.}~\bibnamefont {Ettengruber}},\ and\ \bibinfo {author} {\bibfnamefont {A.}~\bibnamefont {Stuhlfauth}},\ }\href {https://doi.org/10.1103/PhysRevD.109.055046} {\bibfield  {journal} {\bibinfo  {journal} {Phys. Rev. D}\ }\textbf {\bibinfo {volume} {109}},\ \bibinfo {pages} {055046} (\bibinfo {year} {2024})},\ \Eprint {https://arxiv.org/abs/2312.13278} {arXiv:2312.13278} \BibitemShut {NoStop}%
\bibitem [{\citenamefont {Davis}\ and\ \citenamefont {Young}(2017)}]{Davis:2016uyk}%
  \BibitemOpen
  \bibfield  {author} {\bibinfo {author} {\bibfnamefont {E.~D.}\ \bibnamefont {Davis}}\ and\ \bibinfo {author} {\bibfnamefont {A.~R.}\ \bibnamefont {Young}},\ }\href {https://doi.org/10.1103/PhysRevD.95.036004} {\bibfield  {journal} {\bibinfo  {journal} {Phys. Rev. D}\ }\textbf {\bibinfo {volume} {95}},\ \bibinfo {pages} {036004} (\bibinfo {year} {2017})},\ \Eprint {https://arxiv.org/abs/1611.04205} {arXiv:1611.04205 [nucl-ex]} \BibitemShut {NoStop}%
\bibitem [{\citenamefont {Friedman}\ and\ \citenamefont {Gal}(2008{\natexlab{b}})}]{Friedman:2008ef}%
  \BibitemOpen
  \bibfield  {author} {\bibinfo {author} {\bibfnamefont {E.}~\bibnamefont {Friedman}}\ and\ \bibinfo {author} {\bibfnamefont {A.}~\bibnamefont {Gal}},\ }\href@noop {} {\bibfield  {journal} {\bibinfo  {journal} {arXiv preprint}\ } (\bibinfo {year} {2008}{\natexlab{b}})},\ \bibinfo {note} {hep-ph},\ \Eprint {https://arxiv.org/abs/0803.3696} {arXiv:0803.3696} \BibitemShut {NoStop}%
\end{thebibliography}%

\end{document}